\documentclass[nonacm]{acmart}

\AtBeginDocument{%
  \providecommand\BibTeX{{%
    \normalfont B\kern-0.5em{\scshape i\kern-0.25em b}\kern-0.8em\TeX}}}

\setcopyright{none}
\renewcommand\footnotetextcopyrightpermission[1]{}
\usepackage{listings}

\usepackage{xcolor}
\usepackage{mathtools}
\usepackage{subcaption}

\usepackage{caption}

\usepackage[ruled, lined, linesnumbered, commentsnumbered, longend]{algorithm2e}

\SetCommentSty{mycommfont}

\definecolor{codegreen}{rgb}{0,0.6,0}
\definecolor{codegray}{rgb}{0.5,0.5,0.5}
\definecolor{codepurple}{rgb}{0.58,0,0.82}
\definecolor{backcolour}{rgb}{0.95,0.95,0.95}

\definecolor{delim}{RGB}{20,105,176}
\definecolor{numb}{RGB}{106, 109, 32}
\definecolor{string}{rgb}{0.64,0.08,0.08}

\lstdefinestyle{qfaasstyle}{
    backgroundcolor=\color{backcolour},   
    commentstyle=\color{codegreen},
    keywordstyle=\color{magenta},
    numberstyle=\tiny\color{codegray},
    stringstyle=\color{codepurple},
    basicstyle=\ttfamily\footnotesize,
    breakatwhitespace=false,         
    breaklines=true,                 
    captionpos=b,                    
    keepspaces=true,                 
    numbers=left,                    
    numbersep=10pt,                  
    showspaces=false,                
    showstringspaces=false,
    showtabs=false,                  
    tabsize=2,
    frame=single,
    xleftmargin=0.5in,
    xrightmargin=.25in
}

\lstdefinelanguage{json}{
    rulecolor=\color{black},
    postbreak=\raisebox{0ex}[0ex][0ex]{\ensuremath{\color{gray}\hookrightarrow\space}},
    upquote=true,
    morestring=[b]",
    literate=
     *{0}{{{\color{numb}0}}}{1}
      {1}{{{\color{numb}1}}}{1}
      {2}{{{\color{numb}2}}}{1}
      {3}{{{\color{numb}3}}}{1}
      {4}{{{\color{numb}4}}}{1}
      {5}{{{\color{numb}5}}}{1}
      {6}{{{\color{numb}6}}}{1}
      {7}{{{\color{numb}7}}}{1}
      {8}{{{\color{numb}8}}}{1}
      {9}{{{\color{numb}9}}}{1}
      {\{}{{{\color{delim}{\{}}}}{1}
      {\}}{{{\color{delim}{\}}}}}{1}
      {[}{{{\color{delim}{[}}}}{1}
      {]}{{{\color{delim}{]}}}}{1},
}

\newcommand*{\code}{\lstinline[keywordstyle=\color{black}, basicstyle=\ttfamily\small\color{black}]}

\begin{document}

\title{QFaaS: A Serverless Function-as-a-Service Framework for Quantum Computing}

\author{Hoa T. Nguyen}
\email{thanhhoan@student.unimelb.edu.au}
\affiliation{%
  \institution{Cloud Computing and Distributed Systems (CLOUDS) Laboratory, School of Computing and Information Systems, The University of Melbourne}
  \city{Parkville}
  \state{Victoria}
  \country{Australia}}

\author{Muhammad Usman}
\affiliation{%
  \institution{School of Physics, The University of Melbourne}
  \city{Parkville}
  \state{Victoria}
  \country{Australia}}
\email{musman@unimelb.edu.au}

\author{Rajkumar Buyya}
\affiliation{%
  \institution{Cloud Computing and Distributed Systems (CLOUDS) Laboratory, School of Computing and Information Systems, The University of Melbourne}
  \city{Parkville}
  \state{Victoria}
  \country{Australia}}
\email{rbuyya@unimelb.edu.au}

\renewcommand{\shortauthors}{H.T. Nguyen, M. Usman, and R. Buyya}

\begin{abstract}
Recent breakthroughs in quantum hardware are creating opportunities for its use in many applications. However, quantum software engineering is still in its infancy with many challenges, especially dealing with the diversity of quantum programming languages and hardware platforms. To alleviate these challenges, we propose QFaaS, a novel Quantum Function-as-a-Service framework, which leverages the advantages of the serverless model and the state-of-the-art software engineering approaches to advance practical quantum computing. Our framework provides essential components of a quantum serverless platform to simplify the software development and adapt to the quantum cloud computing paradigm, such as combining hybrid quantum-classical computation, containerizing functions, and integrating DevOps features. We design QFaaS as a unified quantum computing framework by supporting well-known quantum languages and software development kits (Qiskit, Q\#, Cirq, and Braket), executing the quantum tasks on multiple simulators and quantum cloud providers (IBM Quantum and Amazon Braket). This paper proposes architectural design, principal components, the life cycle of hybrid quantum-classical function, operation workflow, and implementation of QFaaS. We present two practical use cases and perform the evaluations on quantum computers and simulators to demonstrate our framework's ability to ease the burden on traditional engineers to expedite the ongoing quantum software transition.
\end{abstract}




\maketitle

\section{Introduction}
With rapid advances in quantum computing, it is becoming a critical technology attracting significant investment at the global level. In terms of quantum hardware development, IBM is one of the leading companies with the world's most powerful 127-qubit quantum computer based on superconducting technologies released in 2021 \cite{ibm127qubit}. They also have a promising roadmap to develop a quantum computer with 1,121 qubits by 2023 \cite{ibmroadmap}. Apart from IBM, many other major companies, such as Microsoft, Google, DWave, Rigetti, IonQ, and several research groups worldwide, are also working towards building a large-scale quantum computer with fault-tolerant error correction capabilities. \cite{quantumsurvey-raj}. They strive to make quantum computing trustworthy enough to tackle computationally intractable tasks for classical supercomputers. Therefore, these rapid advancements in quantum hardware trigger more investments in quantum software engineering and quantum algorithms development to maximize the practical use of quantum computers. 

There are now legitimate shreds of evidence that quantum computers can solve many complex problems which are challenging to tackle with classical supercomputers, ranging from chemistry problems \cite{Kandala2017Hardware-efficientMagnets} to machine learning \cite{Biamonte2017QuantumLearning}, cryptography\cite{Quan2021AVerification}, and finances \cite{Griffin2021QuantumFinance}. Some notable algorithms have been proposed in the last few decades, such as Deutsch-Jozsa's \cite{djalgo}, Shor's \cite{shoralgo}, and Grover's \cite{groveralgo}. We have also witnessed highly sophisticated quantum algorithms such as Quantum Approximate Optimization Algorithm (QAOA) \cite{qaoa} and Variational Quantum Eigensolver (VQE) \cite{vqe} in recent years. These algorithms have been directly applied to problems of practical relevance, albeit at the proof-of-concept level due to hardware limitations. In terms of quantum software engineering, the number of new quantum programming languages, software development kits (SDKs), and platforms has been accelerating rapidly.  Currently, a user can develop quantum applications using popular SDKs and languages like Qiskit \cite{qiskit}, Cirq \cite{cirq}, Q\# \cite{qsharp}, and Braket \cite{amz-braket-sdk}. Afterward, those quantum applications can be compiled and run on a quantum simulator or sent to a physical quantum computer via cloud-based services such as IBM Quantum \cite{ibmq}, and Amazon Braket \cite{amazonbraket}. However, quantum software engineering and quantum cloud computing are still confronting many challenges  and some of these are discussed in the next sections.

\subsection{Challenges for quantum software engineering}
Quantum software engineering is a rapidly developing emerging area, and there are many open challenges. First, the development of quantum applications is time-consuming for software engineers, mainly because of the requirement of prior quantum knowledge. Quantum programming is underpinned by the principles of quantum mechanics, which are quite different from the traditional models. Therefore a quantum programmer must overcome the hurdle of learning quantum mechanics to develop applications. A basic example is the difference of the fundamental unit: a classical bit has two states: 0 and 1, whereas a quantum bit (qubit) could also be placed in a \textit{"superposition"} state, i.e., a combination state of 0 and 1 simultaneously \cite{qbook-nielsenchuang}. Second, quantum computing is a promising way to solve several intractable tasks even for a classical supercomputer, but it may not entirely replace classical computers. In other words, there are many tasks in which both quantum and classical approaches could have the same performance, such as performing a simple calculation. While classical solutions are still dominant in today's industry, it is challenging to decide which approach is more suitable when transferring the existing classical systems to quantum: whether to replace the entire system with quantum or integrate quantum into the already well-established classical system \cite{Grossi2021AComputing}. Besides, the variety of current quantum SDKs and the heterogeneous quantum technologies could confuse software engineers in picking an appropriate technique for their software. Each SDK and language has different environment configuration requirements, syntax, and methods to connect with the quantum simulator or quantum computer. Additionally, there is no well-known standardization or life cycle in quantum software engineering similar to practices like Agile and DevOps in the traditional realm \cite{Weder2021QuantumLifecycle}.

Recently, numerous solutions have been proposed to eliminate these burdens and accelerate quantum software development \cite{quantumsurvey-raj}. Many studies have focused on developing new quantum SDKs \cite{qiskit, cirq, Rigetti2021ForestFramework}, quantum programming languages \cite{qsharp, Fu2021Quingo:Features, Mccaskey2021ExtendingComputing}, and platforms \cite{quantumpath}. However, few studies \cite{Dreher2019PrototypeDevelopment} have considered the potential approach of leveraging modern classical techniques and computation models to apply in the quantum realm, which has motivated us to contribute to this research.

\subsection{Challenges for quantum cloud computing}

The most widely adopted way to access today’s quantum computers is through a cloud service from external vendors, such as IBM Quantum \cite{ibmq}, Amazon Braket \cite{amazonbraket}, Azure Quantum \cite{azurequantum}, and Google Quantum Computing Service \cite{googleqcs}. However, the difference between quantum and classical approaches poses many challenges for the quantum cloud computing paradigm. First, we can not permanently deploy a quantum application on a cloud-based quantum computer as a service for invoking many times from the end-users, similar to what we do with traditional ones. Instead, we need to build a suitable quantum circuit for our application, then deploy them to quantum computers, and wait for the execution result \cite{Garcia-Alonso2022QuantumGateway}. Additionally, quantum cloud providers and customers need to establish a win-win paradigm to maximize quantum advantages while optimizing the budget and the resources. The current pay-per-use pricing model offered by cloud vendors such as Amazon Braket \cite{amazonbraket} needs to go along with the computing model like serverless, a trending model for the classical cloud, to balance the benefit of both parties. An example of this approach is the concept of Quantum Serverless \cite{Johnson2021QuantumServerless}, proposed by IBM in their Quantum Summit 2021, coming up with the development of Qiskit Runtime \cite{qiskitruntime}. However, by sticking to a specific quantum vendor and technology, we could encounter another popular challenge known as the \textit{"vendor lock-in"} problem \cite{Hassan2021SurveyComputing}. Therefore, an effort to make a universal quantum serverless platform, working with multiple quantum SDKs and providers, is another pivotal inspiration for our proposed work.

\subsection{Contributions}
In order to address and mitigate the challenges highlighted above, we propose QFaaS \textit{(\textbf{Q}uantum \textbf{F}unction-\textbf{a}s-\textbf{a}-\textbf{S}ervice)}, a universal quantum serverless framework, which offers the function-as-a-service deployment model for quantum computing. Our framework could ease the quantum software development process, enabling traditional software engineers to quickly adapt to the quantum transition while continuously utilizing their familiar models and techniques.

The key contributions of our proposed research are as follows:
\begin{itemize}
    \item We design a novel framework for developing quantum function as a service, supporting popular quantum SDKs and languages, including Qiskit, Cirq, Q\#, and Braket, to perform the computation on classical computers, quantum simulators, and quantum computers provided by multiple vendors (IBM Quantum and Amazon Braket).
    \item We evaluate the suitability, conduct empirical investigations, and apply state-of-the-art classical technologies and models, such as containerization, GitOps, and function-as-a-service for quantum software engineering. By leveraging the Docker container with Kubernetes as the underlying technique, our framework is portable and scalable for further migration or expansion to a large-scale system.
    \item We utilize the DevOps techniques in operating QFaaS, including continuous integration and continuous deployment, which supports to automate the quantum software development cycle, from quantum environment setup to hybrid quantum-classical function deployment.
    \item We introduce a unified 6-stage life cycle for a quantum function, from function development, deployment, pre-processing, backend selection, quantum execution, and post-processing. This lifecycle provides a baseline for a quantum software engineer to plan and organize their software development process.
    \item We propose two operation workflows for both kinds of users: quantum software engineers and end-users, to utilize our framework for the hybrid quantum-classical applications. The end-users can access the deployed function as a service through the QFaaS API (Application Programming Interface) gateway. Our framework also provides multiple ways for users to interact with the core components, including QFaaS Dashboard (a modern web-based application), QFaaS CLI (an interactive command-line tool), and QFaaS Core APIs.
    \item We have implemented two application use cases with QFaaS to validate our proposed design and demonstrate how our framework can facilitate quantum software development in practice. We also conduct a set of benchmark tests to evaluate the performance of our framework and offer an insight into the current status of today's quantum computers and simulators. This paper proposes the framework and essential implementation, but we have a viable plan to add additional functionality to the QFaaS platform. Ultimately, our framework is expected to be a universal environment for designing advanced practical quantum-classical applications.
\end{itemize}

The rest of the paper is organized as follows: After introducing the fundamentals of quantum computing, section 2 presents the current state of quantum software development, quantum computing as a service (QCaaS) model, and serverless quantum computing. Section 3 discusses the related work and briefly compares our framework's benefits with existing work. Section 4 introduces the details of the QFaaS framework, including the design principle, principal components, structure and life cycle of a quantum function, and the operation workflow of our framework. Section 5 describes the design and implementation of QFaaS core components and functions. Then, section 6 demonstrates the operation of QFaaS in two use cases and its performance. Following the discussion of the advantages of our framework for software engineering in section 7, we conclude and present our plan the future work in section 8.

\section{Background}
\subsection{The Fundamentals of Quantum Computing}
This section briefly summarizes several essential characteristics and building blocks of gate-based quantum computing before diving into the state-of-the-art development of quantum software engineering and serverless quantum computing.

\subsubsection{Qubits, Superposition and Entanglement}

Quantum computing is based on the theory of quantum mechanics and, therefore, is fundamentally different from classical computing \cite{qbook-nielsenchuang}. The basic units of classical and quantum computing are strikingly different at the fundamental level: a classical bit and a quantum bit (or qubit). A bit has two states for computation, either 0 or 1. Besides these classical states, a qubit can have a \textit{superposition} state, i.e., a combination of states 0 and 1 simultaneously. Often quantum algorithms can achieve exponential speed-up by leveraging this characteristic compared with the classical solution.
We can describe the general state of a qubit $|\psi\rangle$ as follows:
\begin{displaymath}
    |\psi\rangle = \alpha|0\rangle + \beta|1\rangle
\end{displaymath}
where $\alpha, \beta \in \mathbb{C}$ are complex numbers.
However, whenever we measure the superposition state, it could collapse to one of the classical states (i.e., 0 or 1): 
\begin{displaymath}
    ||\alpha||^2 + ||\beta||^2 = 1
\end{displaymath}
where $||\alpha||^2$ and $||\beta||^2$ is the probability of 0 and 1 as a result after measuring qubit $|\psi\rangle$. Hence, it is not straightforward to design a useful quantum algorithm by only utilizing the superposition attribute. 

Another critical characteristic of qubits that could be leveraged to design quantum algorithms is \textit{entanglement}. Entanglement is a robust correlation between two qubits, i.e., one party always knows precisely the state of the other, even if they are very far away. In other words, if a pure state  $|\psi\rangle_{AB}$  on two systems A and B cannot be written as $|\psi\rangle_A \otimes|\phi\rangle_B$, we called it is \textit{entangled} \cite{IBMQuantum2022QiskitOnline}.

\subsubsection{Quantum Gates}
To perform the quantum operations on qubits, we apply quantum gates, which is conceptually similar to how we apply classical gates, such as AND, OR, XOR, and NOT on classical bits to perform classical computation. The quantum gates are always reversible, i.e., a qubit does not change its state if we apply the same quantum gate twice between the qubit initialization and the measurement. For example, we could use the Hadamard (H) gate on qubit $|0\rangle$ to create an equal superposition $|+\rangle = \tfrac{1}{\sqrt{2}}(|0\rangle+|1\rangle)$. If we apply the H gate again, the $|+\rangle$ state will be reversed to the original $|0\rangle$  state. A quantum gate U could be represented by a unitary matric such that $U^\dagger U = \mathbb{I}$  where I is the identity matrix. The general representation of a single-qubit gate U is as follows:

\begin{displaymath}
U(\theta, \phi, \lambda) = \begin{bmatrix} \cos(\theta/2) & -e^{i\lambda}\sin(\theta/2) \\
            e^{i\phi}\sin(\theta/2) & e^{i\lambda+i\phi}\cos(\theta/2)
     \end{bmatrix}
\end{displaymath}
where $\theta, \phi, \lambda$ are the different parameters for each specific gate \cite{IBMQuantum2022QiskitOnline}. 

We can categorize quantum gates into two main types: single-qubit gates and multiple-qubit gates. Some popular single-qubit gates are Pauli gates (Pauli-X, Pauli-Y, Pauli-Z), the Hadamard (H) gate, and the Phase (P) gate. For example, the Hadamard (H) gate could be represented as the following (with $\theta, \phi, \lambda$ = $\tfrac{\pi}{2}, 0, \pi$, respectively):
\begin{displaymath}
H = U(\tfrac{\pi}{2}, 0, \pi) = \tfrac{1}{\sqrt{2}}\begin{bmatrix} 1 & 1 \\
            1 & -1
     \end{bmatrix}
\end{displaymath}

We can also apply quantum gates to multiple qubits simultaneously by using multi-qubit gates, such as Controlled-NOT (CNOT) gate and Toffoli gate. CNOT gate, for instance, is a controlled two-qubit gate, i.e., the target qubit will change its state if the control qubit is $|1\rangle$  and will not change its state if the control qubit is $|0\rangle$ \cite{IBMQuantum2022QiskitOnline}. The matrix representation of the CNOT gates is as follows: 
\begin{displaymath}
\text{CNOT} = \begin{bmatrix} 1 & 0 & 0 & 0 \\
                              0 & 0 & 0 & 1 \\
                              0 & 0 & 1 & 0 \\
                              0 & 1 & 0 & 0 \\
              \end{bmatrix}
\end{displaymath}
We can create the \textit{entangled} state (Bell state) by applying the CNOT gate to $|0+\rangle$ state:

\begin{displaymath}
\text{CNOT}|0{+}\rangle = \text{CNOT}( \tfrac{1}{\sqrt{2}}(|00\rangle + |01\rangle))
 = \tfrac{1}{\sqrt{2}}(|00\rangle + |11\rangle)
\end{displaymath}

\subsubsection{Quantum Circuits and Quantum Algorithms}
When implementing a quantum algorithm using the gate-based approach, we need to connect an appropriate combination of quantum gates to build quantum circuits. A quantum circuit's general operation includes three main stages: 1) Initializing the qubits, 2) Applying the quantum gates, and 3) Performing the measurement. 

\begin{figure}[htbp] 
    \centering 
    \includegraphics[scale=0.26]{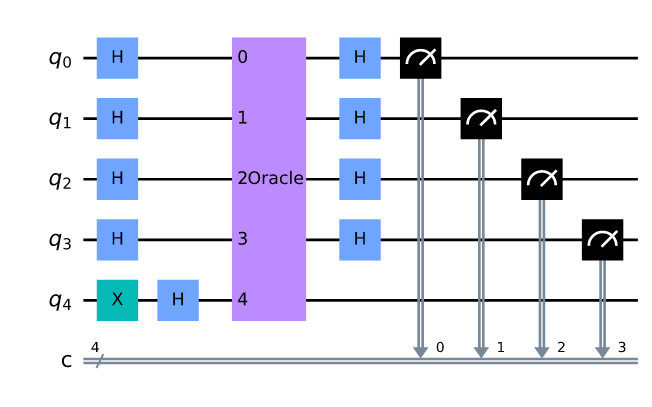} 
    \caption{An example quantum circuit for the Deutsch-Jozsa Algorithm (generated by using Qiskit)} 
    \label{fig:djalgo} 
\end{figure}

For example, the quantum circuit shown in Figure \ref{fig:djalgo} implements Deutsch-Jozsa's algorithm \cite{djalgo}. The main objective of this algorithm is to determine whether the property of the oracle is constant (i.e., always return 0 or 1) or balanced (i.e., return 0 and 1 with the same probability). The oracle in this circuit is a \textit{"black box"} where we do not know which binary value is inside. However, when we query it with arbitrary input data, it will return a binary answer, either 0 or 1. For the traditional approach, we need to interact with the oracle at least two times and at most $2^{n-1}+1 $ times, where n is the number of input bits. Using the Deutsch-Jozsa algorithm, we need to query the oracle only once to get the final result. If the measurement outcomes of all the qubits are 0, we can determine that the oracle is constant; otherwise, it is balanced. This algorithm was also the first to demonstrate that quantum computers could outperform the classical computer in 1992. \cite{IBMQuantum2022QiskitOnline}.

\subsection{Quantum Software Engineering}
\subsubsection{Quantum SDKs and languages}
At the moment, when starting to develop a new quantum application, we do not necessarily have to learn a new programming language. Instead, we can continue utilizing our familiar languages, such as Python \cite{qiskit, cirq} or C++ \cite{Mccaskey2021ExtendingComputing}. Fortunately, we have a ton of available quantum software development kits (SDKs) and programming languages to choose from, thanks to the productive work in this field of both the quantum industry and the research community. Some popular SDKs and languages that were originated from well-known companies are:
\begin{itemize}
    \item \textbf{Qiskit} \cite{qiskit} is well-known and probably the most popular (as per the Github repository’s star count) Python-based open-source SDK for developing gate-based quantum programs, initially developed by IBM. Qiskit has a wide range of additional libraries and support tools and is best suited to the IBM Quantum Cloud platform \cite{ibmq}. We can create a Qiskit program by multiple methods, by writing the Python script, using Jupyter Notebook, or using an online Quantum Composer provided by IBM Quantum. Then, we can run that program by using a built-in simulator (such as Aer Simulator, or QASM simulator) or sending it to execute in an IBM quantum computer.
    \item \textbf{Cirq} \cite{cirq} is another prevalent open-source Python software library introduced by Google for quantum computing. This SDK supports us in writing, manipulating, and optimizing quantum gate-based circuits. Cirq programs can run on built-in simulators (wave functions and density matrices), and Google’s quantum processors \cite{googleqcs}. It is also an underlying SDK for TensorFlow Quantum \cite{tensorflowquantum}, a high-level library for performing hybrid quantum-classical machine learning tasks.
    \item \textbf{Q\#} \cite{qsharp} is a new programming language from Microsoft for developing and executing quantum algorithms. It comes along with Microsoft’s Quantum Development Kit, which includes a set of toolkits and libraries for quantum software development. Q\# also offers many ways to use, by creating standalone programs, using Jupyter notebooks, using the command line, or integrating with host languages such as C\# or Python.
    \item \textbf{Braket} \cite{amz-braket-sdk} is an emerging Python-based SDK of Amazon to interact with their quantum computing service, named Amazon Braket \cite{amazonbraket}. This SDK provides multiple ways to prototype and develop hybrid quantum applications, then run them on simulators (fully managed and local simulators) or quantum computers (provided by third-party hardware companies D-Wave, Rigetti, and IonQ).
    
\end{itemize}
Besides, there are numerous quantum languages and SDKs proposed by research groups and other companies over the world, such as \textit{Forest} and \textit{pyQuil} by Rigetti \cite{Rigetti2021ForestFramework}; \textit{Strawberry Fields} \cite{Killoran2019StrawberryComputing} and \textit{PennyLane} \cite{Bergholm2020PennyLane:Computations} by Xanadu;  \textit{Quingo} \cite{Fu2021Quingo:Features}; \textit{QIRO} \cite{Ittah2022QIRO:Optimization}; or \textit{qcor} \cite{qcor}.

\subsubsection{NISQ era and the Hybrid Quantum-Classical model}
John Preskill proposed the \textit{“Noisy Intermediate-Scale Quantum (NISQ)”} term in 2018 \cite{Preskill2018QuantumBeyond} to describe the current state of quantum computers. This term indicates two characteristics of today’s quantum devices, including \textit{“noisy,”} i.e., unstable and error-prone quantum state due to the affection of various environmental actions, and \textit{“intermediate scale,”} i.e., the quantum volume is at the medium level, with about a few tens of qubits \cite{Weder2020TheLifecycle}. The most powerful quantum computer with 127 qubits, Eagle (released by IBM in 2021 \cite{ibm127qubit}) could also be categorized as a NISQ computer. 

These limitations of current quantum devices pose many challenges for quantum software engineers to develop, execute, and optimize quantum applications.
Due to the NISQ nature, the typical pattern for developing today’s quantum programs combines quantum and classical parts \cite{Leymann2021HybridPerspective}. In this hybrid model, the classical components are mainly used for pre-processing and post-processing the data. In contrast, the remaining part is sent to quantum computers for computation. The quantum execution parts are repeated many times and measure the average values to mitigate the error caused by the noisy quantum environment. An example of the hybrid quantum-classical model is Shor’s algorithm \cite{shoralgo} to find prime factors of integer numbers. In this algorithm, we execute the period-finding part, leveraging the Quantum Fourier Transform on quantum computers and then performing the classical post-process to measure the prime factors based on the outcome of the quantum computation part. Other hybrid computation examples are the Quantum Approximate Optimization Algorithm (QAOA) \cite{qaoa}, or the Variational Quantum Eigensolver (VQE) \cite{vqe}.

\subsubsection{Variation of Quantum Software Development Lifecycle}
In traditional software development, a lifecycle is an overall procedure of developing and operating an application, involving many steps from designing, executing, maintaining, investigating, and adapting software \cite{Weder2020TheLifecycle}. Standardizing a lifecycle for quantum software development is also inevitably essential to ensure stability and scalability for the long term. Several studies have proposed various software lifecycles for quantum computing recently.

In recent years, DevOps (Development and Operations) has been a trending model adopted by numerous companies to accelerate the software development process and increase revenue faster. DevOps is a union of people, processes, and products whose primary goal is to deliver value to end-users continuously \cite{devops}. Gheorghe-Pop et al. \cite{Gheorghe-Pop2020QuantumComputing} proposed the Quantum DevOps workflow for extending traditional DevOps phases into quantum software engineering. This workflow includes six continuous steps in each Dev and Ops phase: 1) Plan $ \rightarrow $ 2) Code $ \rightarrow $ 3) Build $ \rightarrow $ 4) Test $ \rightarrow $ 5) Release $ \rightarrow $ 6) Feedback. Benjamin et al. \cite{Weder2020TheLifecycle} proposed an overall 10-step quantum software lifecycle. These steps include: 1) Quantum-Classical Splitting $ \rightarrow $ 2) Hardware-independent implementation $ \rightarrow $ 3) Quantum Circuit Enrichment $ \rightarrow $ 4) Hardware-Independent Optimization $ \rightarrow $ 5) Quantum Hardware Selection $ \rightarrow $ 6) Readout-Error Mitigation Preparation $ \rightarrow $ 7) Compilation and Hardware dependent Optimization $ \rightarrow $ 8) Integration $ \rightarrow $ 9) Execution $ \rightarrow $ 10) Result Analysis. Then, they proposed an altered lifecycle \cite{Weder2021QuantumLifecycle} with eight steps, including 1) Requirement Analysis $ \rightarrow $ 2) Quantum-Classical Splitting $ \rightarrow $ 3) Architecture and Design $ \rightarrow $ 4) Implementation $ \rightarrow $ 5) Testing $ \rightarrow $ 6) Deployment $ \rightarrow $ 7) Observability $ \rightarrow $ 8) Analysis. Along with the Quingo framework proposed in \cite{Fu2021Quingo:Features}, the authors also suggested a six-phase life cycle for a quantum program, including 1) Editing $ \rightarrow $ 2) Classical compiling $ \rightarrow $ 3) Pre-executing $ \rightarrow $ 4) Quantum compiling $ \rightarrow $ 5) Quantum executing $ \rightarrow $ 6) Classical post-executing.

Due to the immature development and lack of standardization, more efforts are still needed to advance this field and adapt to quantum hardware's continuous growth. From the practical point of view, in the design and implementation of the QFaaS framework, we customized and proposed a sample 6-stage lifecycle for quantum function development, which will be described in detail in section \ref{qfaaslifecycle}.

\subsection{Quantum Computing as a Service (QCaaS)}
Today's quantum computers are made available to the industry and research community as a cloud service by a quantum cloud provider \cite{Garcia-Alonso2022QuantumGateway}. This scheme is well known as Quantum Computing as a Service (QCaaS or QaaS), which corresponds with well-known paradigms in cloud computing such as Platform as a Service (PaaS) or Infrastructure as a Service (IaaS) \cite{cloud-raj}. In terms of QCaaS, software engineers can develop quantum programs and send them to quantum cloud providers to execute that program on appropriate hardware. After finishing the computation, the users only need to pay for the actual execution time of the quantum program (pay-per-use model). In this way, QCaaS is an efficient way that optimizes the user's budget for using quantum computing services and the provider's resources.

Many popular cloud providers nowadays offer quantum computing services using their quantum hardware, such as IBM Quantum \cite{ibmq}, which is publicly accessible for everyone in their early phase. Besides, other quantum computing services (such as Amazon Braket \cite{amazonbraket}, and Azure Quantum \cite{azurequantum}) collaborate with other hardware companies such as D-Wave, Rigetti, and IonQ to provide commercial services. For example, Amazon Braket, a new Quantum Computing service of Amazon Web Services (AWS), currently offers the pay-per-use pricing model as Table \ref{tab:braket}:

\begin{table}[h!]
  \caption{Amazon Braket Pricing for using Quantum Computers (April, 2022) \cite{amazonbraket}}
  \label{tab:braket}
  \begin{tabular}{cccl}
    \toprule
    \textbf{Hardware Provider} & \textbf{QPU Family} & \textbf{Per-task price} (\$) & \textbf{Per-shot price}   (\$) \\
    \midrule
    D-Wave & 2000Q & 0.3 & 0.00019 \\
D-Wave & Advantage & 0.3 & 0.00019 \\
IonQ & IonQ device & 0.3 & 0.01 \\
OQC & Lucy & 0.3 & 0.00035 \\
Rigetti & Aspen-11 & 0.3 & 0.00035 \\
Rigetti & M-1 & 0.3 & 0.00035 \\
  \bottomrule
\end{tabular}
\end{table}

By running a gate-based quantum application on the Rigetti M-1 quantum computer with 10,000 shots (i.e., iterated execution) at Amazon Braket, we need to pay:

\textit{Total charges = Task charge + Shots charge}

\textit{= the number of task * per-task price + the number of shots * per-shot price }

= 0.3*1 + 10,000 * 0.00035 = \$3.80 \cite{amazonbraket}.

However, this paradigm still faces many challenges before solving real-world applications due to the limitation of today’s NISQ computers \cite{Preskill2018QuantumBeyond}. These devices have a small number of qubits that are error-prone and limited in capabilities. Therefore, improving the quality and quantity of qubits for quantum computers will accelerate of QCaaS model and quantum software development.

\subsection{Serverless Quantum Computing}
\subsubsection{Serverless Computing and Function as a Service (FaaS)}
In the classical computing domain, serverless is an emerging model and could be considered a second phase for traditional cloud computing \cite{Schleier-Smith2021WhatBecome}. Serverless does not mean the absence of physical servers; it refers to an execution model that simplifies the application development without worrying about setting up the underlying system infrastructure. In other words, serverless implies that the existence of the servers is abstracted away from the software engineers. This computing model fits with modern software architecture, especially the microservice applications, where the overall application is decomposed into multiple small and independent modules \cite{Eismann2021ServerlessHow}. The serverless computing concept generally incorporates both Function-as-a-Service (FaaS) and Backend-as-a-Service (BaaS) models \cite{Taibi2021ServerlessHeading}. FaaS refers to the stateless ephemeral function model where a function is a small and single-purpose artifact with few lines of programming code. BaaS is a concept to describe serverless-based file storage, database, streaming, and authentication service.

As FaaS is a subset of the serverless model, its main objective is to provide a concrete and straightforward way to implement software compared with traditional monolith architecture. FaaS allows the software engineer to focus only on coding rather than environmental setup and infrastructure deployment.  A function can be triggered by a database, object storage, or deployed as a REST API and accessed via an HTTP connection. Functions also need to be scalable, i.e., automatically scaling down when idle and scaling up when the request demand increases. In this way, a FaaS platform could be an efficient way to optimize the resource for providers and reduce costs for customers. There are numerous open-source FaaS platforms in the traditional cloud-native landscape, such as OpenFaaS, OpenWhisk, Kubeless, Knative, and also many commercial serverless platforms such as AWS Lambda, Azure Functions, Google Cloud Functions \cite{Hassan2021SurveyComputing}.

\subsubsection{Severless Quantum Computing}
In November 2021, along with the 127-qubit quantum computer, IBM also introduced the concept of Quantum Serverless \cite{Johnson2021QuantumServerless}, an adapted execution model for combing quantum and classical resources together. This model followed the principles of the traditional serverless platform, which embodies four key characteristics: 1) The only job of software engineers is to focus on their coding without any concern about infrastructure management; 2) All components are cloud-based services; 3) The services are scalable and 4) It fits with the pay-per-use pricing model. IBM also introduced Qiskit Runtime \cite{qiskitruntime} in 2021, which allows users to execute the pre-built circuit by their developer teams, and it could be a premature example of their proposed concept.

A serverless quantum computing model could also be a viable solution for utilizing today’s quantum computers effectively. Indeed, by decomposing a monolith application into multiple single-purpose functions, we could distribute them to various backend devices. This approach is well-suited to the current state of NISQ devices, in which each device has limited resources and could be accessed anywhere through the quantum cloud. Besides, we could implement a hybrid quantum-classical model by combining quantum functions and classical functions in a unified application. This approach could leverage the power of existing quantum computers to facilitate new promising techniques, such as hybrid quantum-classical machine learning \cite{Biamonte2017QuantumLearning}.
Although we can adopt the serverless model for quantum computing, the way a traditional service and quantum service can be deployed and executed are different. We can deploy a service directly and permanently to a classical server one time, and it could be invoked many times by the end-users later on. However, we cannot do the same thing with quantum computers, i.e., a quantum program cannot be deployed persistently in a specific quantum computer \cite{Garcia-Alonso2022QuantumGateway}. With today’s quantum computer, an appropriate quantum circuit (based on user inputs) needs to be built every time we execute a specific task. Then, that circuit will be transpiled to corresponding quantum system-level languages (such as QASM \cite{qasm}) before being sent to a quantum cloud service for execution. Therefore, an adaptable serverless model for executing quantum tasks is needed to address this challenge.

By leveraging the ideas of the serverless model and combining quantum and classical parts in a single service, we can adapt to the current nature of quantum cloud services, accelerate the software development process and optimize quantum resource consumption. This kind of computing model could be a potential approach to enable quantum software engineers and end-users can realize the actual advantages of quantum computing and explore more complicated quantum computation in the future.

\section{Related Work}
This section discusses the related work in the context of frameworks for developing quantum applications. Table \ref{tab:relatedwork} summarizes the difference between QFaaS and related platforms in the context of various capabilities offered by them.

\begin{table*}[htbp]
  \caption{Feature comparison of QFaaS and Related Work for Quantum Software Development \textit{(N/A: No information available)}}
  \label{tab:relatedwork}
  \centering
  \renewcommand{\arraystretch}{1.2}
  \begin{tabular}{p{0.18\textwidth}p{0.08\textwidth}p{0.1\textwidth}p{0.12\textwidth}p{0.08\textwidth}p{0.06\textwidth}p{0.08\textwidth}p{0.09\textwidth}}
    \toprule
    \textbf{Features}  & \textbf{algo2qpu} \cite{Sim2018AComputers} & \textbf{SQC} \cite{Strangeworks2022StrangeworksPlatform} & \textbf{QuantumPath} \cite{quantumpath} & \textbf{SCIQC} \cite{Grossi2021AComputing} & \textbf{QAPI} \cite{Garcia-Alonso2022QuantumGateway} & \textbf{Quingo} \cite{Fu2021Quingo:Features} & \textbf{QFaaS} \textit{(Proposed)} \\
    \midrule
    Quantum SDKs and languages  & Forest & Qiskit, Q\#, Cirq, Forest, ProjectQ, ... & Qiskit, Q\#, Cirq, Ocean, pyQuil & Qiskit & N/A & Quingo & Qiskit, Q\#, Cirq, Braket \\
Code Development Environment  & N/A & Web IDE & Visual Editor & N/A & $\times$ & N/A & Web IDE, Local IDE\\
Templates library   & \checkmark & \checkmark  & \checkmark  & $\times$ & $\times$ & $\times$ & \checkmark \\
Quantum + Classical integration  & \checkmark & \checkmark & \checkmark & \checkmark & $\times$ & \checkmark & \checkmark \\
API Gateway  & $\times$ & $\times$ & \checkmark  & $\times$ & \checkmark & $\times$ & \checkmark \\
Built-in REST API & $\times$ & \checkmark &  $\times$ & $\times$ & $\times$ & $\times$ & \checkmark \\
Serverless FaaS  & $\times$ & $\times$ & $\times$ & \checkmark & $\times$ & $\times$ & \checkmark\\
Containerization  & $\times$ & $\times$ & $\times$ & \checkmark & $\times$ & $\times$ & \checkmark \\
DevOps \textit{(CI/CD)}  & $\times$ & $\times$ & $\times$ & $\times$ & $\times$ & $\times$ & \checkmark \\
UI Dashboard  & $\times$ & \checkmark & \checkmark & \checkmark & $\times$ & $\times$ & \checkmark \\
CLI tool  & $\times$ & $\times$ & $\times$ & $\times$ & $\times$ & $\times$ & \checkmark \\
Implementation \textit{(with practical use cases)}   & \checkmark   & \checkmark  & \checkmark & $\times$ & \checkmark & \checkmark & \checkmark \\
Job monitoring  & $\times$ & $\times$ & $\times$ & \checkmark & $\times$ & $\times$  & \checkmark\\
Scalability  & $\times$ & $\times$ & $\times$ & N/A & $\times$ & $\times$ & \checkmark \\
Quantum Simulators  & N/A & \checkmark  & \checkmark   & \checkmark   & N/A & \checkmark  & \checkmark \\
Quantum Providers \textit{(S: single, M: multiple)}   & \checkmark (S)  & \checkmark (M)   & \checkmark (M)   & \checkmark (S)   & \checkmark (S)  & N/A & \checkmark (M) \\
  \bottomrule
\end{tabular}
\end{table*}

To the best of our knowledge, existing quantum computing platforms lack various capabilities to provide a universal environment for developing service-based quantum applications.
In 2018, Sim et al. \cite{Sim2018AComputers} proposed \textit{algo2qpu}, a hardware and software agnostic framework that supports designing and testing adaptive hybrid quantum-classical algorithms on the Rigetti cloud-based quantum computer. They implemented two applications in quantum chemistry and quantum machine learning using their proposed framework. This work motivated us to develop a more efficient and flexible framework for hybrid quantum-classical software development.

Strangeworks Quantum Computing (SQC) \cite{Strangeworks2022StrangeworksPlatform} is an online platform that allows us to use various templates, such as Qiskit, Q\#, Cirq, ProjectQ, Ocean, Pennylane, and Forest, to develop quantum programs (using Jupyter Notebook and Python script). These programs can run on IBM Quantum machines and other enterprise quantum hardware. SQC could be a convenient tool for quantum experiments and visualization. However, the running model of SQC and QFaaS is different. SQC provides a \textit{“code-and-run”} environment, where we can write our standalone program and run it directly. In contrast, QFaaS provides a function-as-a-service model, where each quantum function is deployed as a service for end-users through an API gateway. This way, we can easily integrate these quantum services into our existing application, following the microservice application scheme.

Another related work applying the serverless model for quantum computing is proposed by Grossi et al. \cite{Grossi2021AComputing}. In this work, they proposed an architecture design to integrate a Minimum Viable Product (MVP) solution for merging quantum-classical applications with reusable code. This framework leverages multiple cloud-native technologies provided by IBM Cloud, such as IBM Cloud Functions and IBM Containers, to implement Qiskit programs on IBM Quantum. However, the implementation of this proposed system is not available, and it depends solely on IBM-based platforms, which could lead to the vendor lock-in problem of serverless computing.

Garcia-Alonso et al. \cite{Garcia-Alonso2022QuantumGateway} proposed the proof of concept about Quantum API Gateway and provided a simple validation using Python and Flask platform on the Amazon Braket platform. QuantumPath, proposed by Hevia et al. \cite{quantumpath}, is a quantum software development platform aiming to support multiple quantum SDKs and languages, supporting gate-based and annealing quantum applications, and provide multiple design tools for creating a quantum algorithm. QuantumPath does not support serverless and scalability features for further expansion, and it is still in the preliminary phase without providing performance evaluation to validate the proposed design. X.Fu et al. \cite{Fu2021Quingo:Features} proposed the overall framework for developing heterogeneous quantum-classical applications, adapting with NISQ devices. Instead of working with popular quantum languages and SDKs, they also proposed a new programming language, called Quingo, to describe the quantum kernel. Although this is an exciting direction for further quantum framework development, it can face many challenges when developing a new programming language compared to improving the well-known languages. For example, expanding the support for large developer communities, security testing for potential vulnerabilities, and covering all aspects of the quantum and classical computation.

Considering the limitations of the existing platforms, we propose a unified framework for quantum computing that bridges the existing gaps. We focus on building the QFaaS framework by leveraging the advantages of traditional software engineering to quantum computing to alleviate the challenges when developing a hybrid quantum-classical function as a service. Our framework adapts to the NISQ era and is ready for a more stable quantum generation in the near future.

\section{Overview of QFaaS Framework}
This section presents the design principles, main components, function development life cycle, and the operational workflows of QFaaS for developing and using hybrid quantum-classical functions.
\subsection{Design Principles}
We design the QFaaS framework based on the following main principles:
\begin{itemize}
    \item \textbf{Modularity}: The whole framework is built as a combination of multiple modules, where each module manages a specific functionality, thus simplifying further expansion and maintenance. This principle is inspired by the microservice software architecture and \textit{“everything-as-a-service”} (XaaS) paradigm \cite{xaas}. Therefore, we could easily integrate new functionalities into the current framework without affecting other modules.
    \item \textbf{Serverless}: The quantum software engineers only need to focus on their programming, and the framework automatically carries out the rest of the deployment and execution procedures. Once deployed, the end-users can access the deployed services through the cloud-based API gateway in multiple methods. If a pricing strategy is established, they only need to pay for their actual resource usage.
    \item \textbf{Flexibility}: The framework allows users to choose their preferred quantum languages, libraries, and quantum providers to avoid potential vendor lock-in situations. The proposed framework supports the current NISQ computers and multiple quantum simulators. Its architecture provides the flexibility to implement possible extensions to support various quantum technologies in the future.
    \item \textbf{Seamlessness}: The framework supports continuous integration and continuous deployment (CI/CD), which are two of the most essential characteristics of the DevOps life cycle to continuously deliver value to end-users without any interruption. Utilization of this model boosts application development and becomes more reliable when compared with the traditional paradigm \cite{Gheorghe-Pop2020QuantumComputing}.
    \item \textbf{Reliability}: The framework architecture is implemented using state-of-the art technologies to ensure high availability, security, fault tolerance, and trustworthiness of the overall system.
    \item \textbf{Scalability}: As one of the critical characteristics of the serverless model, the framework is scalable and adapts to the actual user requests to optimize both the performance and the resource consumption, eventually providing the optimum cost for end-users.
    \item \textbf{Transparency}: The operation workflow of the framework is transparent to both the quantum software engineers and end-users. It also provides information of diagnosing, troubleshooting, logging, and monitoring for further investigations.
    \item \textbf{Security}: The interactions among different components need to be secure. The framework has in-built identity and access management features to guarantee they have sufficient privileges before performing each operation in the system. 
\end{itemize}

\subsection{QFaaS Architectural Components}
\begin{figure}[htbp]
    \centering
    \includegraphics[scale=0.8]{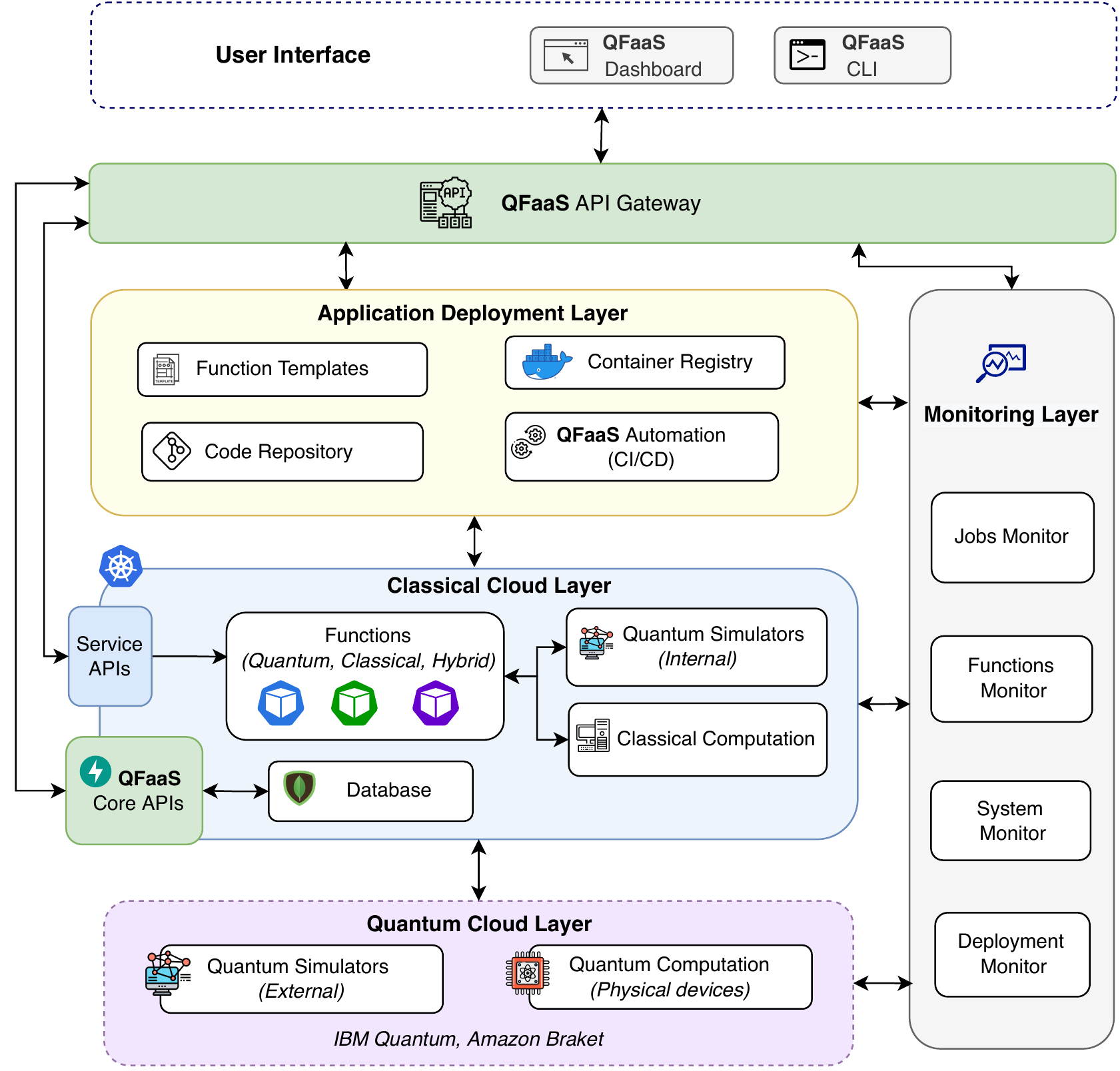}
    \caption{Overview of QFaaS Components and Architecture Design}
    \label{fig:qfass-design}
\end{figure}

The architecture design of QFaaS comprises a set of principal components, which are pluggable and extendable. We break down QFaaS into six components: the QFaaS Core APIs and API Gateway, the Application Deployment Layer, the Classical Cloud Layer, the Quantum Cloud Layer, the Monitoring Layer, and the User Interface. Figure \ref{fig:qfass-design} illustrates the overall design, including the architecture and principal components of our framework.

\subsubsection{QFaaS APIs and API Gateway} QFaaS comprises two kinds of APIs that expose to the appropriate user, called Service APIs and QFaaS Core APIs.
\begin{itemize}
    \item \textbf{Service APIs} are the set of APIs corresponding to the deployed functions. Each function running on the Kubernetes cluster has a unique API URL accessible to an authorized end-user.
    \item \textbf{QFaaS Core APIs} set is one of the most important components in the QFaaS framework. It comprises a set of secure REST APIs, which provide primary operations among all components of the whole system. These APIs simplify the function development, invocation, management and monitoring. QFaaS Core APIs also facilitate the main functionalities of the QFaaS Dashboard and QFaaS CLI tool. We explain the detailed design and implementation of QFaaS Core APIs in Section \ref{section-coreapi}.
\end{itemize}
The API gateway serves as an entrance where users can interact with other components. This gateway routes users’ requests to suitable components and delivers the result back to the users after completing the processing.

\subsubsection{Application Deployment Layer}

This layer serves as a bridge between quantum software engineers and the cloud layers to deploy and publish each function as a service for end-users. It takes the principal responsibility for developing, storing, and deploying quantum functions by combining four key components:

\begin{itemize}
    \item \textbf{Code Repository} is a Git-based platform to store function codes with version control management, which is essential for collaboration in software development. We could deploy this repository privately or publicly with popular Git-based open-source platforms such as Gitlab, Github, and Bitbucket. 
    \item \textbf{Function Templates} include container-based quantum software environment setup and sample function format for all supported quantum SDKs and languages, including Qiskit, Cirq, Q\#, and Braket.
    \item \textbf{QFaaS Automation} implements the Continuous Integration and Continuous Deployment (CI/CD) process, following the DevOps paradigm to ensure the continuous delivery of reliable quantum functions for end-users.
    
    \item \textbf{Container Registry} stores Docker images of functions and environmental setup after implementing the QFaaS Automation process. These images are immutable and could be used to deploy a function as a container-based service to the classical cloud layer.
\end{itemize}

\subsubsection{Classical Cloud Layer}
This layer comprises a cluster of cloud-based classical computers (physical servers or virtual machines), where the QFaaS functions are deployed and executed. All the classical processing tasks, including pre-processing and post-processing, are performed at this layer.

We employed Kubernetes to orchestrate all the \textit{pods} (the container-based unit of Kubernetes) for the QFaaS function across all cluster nodes. Each function will be run on a pod and could be scaled up horizontally by replicating the original pod. All quantum SDKs and languages have their built-in quantum simulator, which can run directly inside a pod at the Kubernetes cluster. We call this kind of simulator as the \textit{internal quantum simulator}, while the \textit{external quantum simulator} is used to indicate the simulator provided by quantum cloud providers.
We also deployed a NoSQL database (MongoDB) on this layer to store the processed job results and information of users, functions, and backends.

\subsubsection{Quantum Cloud Layer}
This layer is the external part, indicating the quantum cloud providers (such as IBM Quantum and Amazon Braket), where the quantum job can be executed in a physical quantum computer.
Quantum providers can provide either quantum simulators or actual quantum computers through their cloud services and could be accessed from the Classical Cloud layer.

\subsubsection{Monitoring Layer}
This layer includes different components which periodically check the status of other layers:
\begin{itemize}
    \item \textbf{Quantum job monitoring} component periodically checks the job status, queuing information, and quantum job result from external quantum providers.
    \item \textbf{Function monitoring} component provides the current status of deployed quantum functions and function usages (such as the number of invocations).
    \item \textbf{System monitoring} component provides the status of the overall systems, such as network status and resource consumption at the Kubernetes cluster.
    \item \textbf{Deployment monitoring} component tracks the function deployment process to provide the quantum software engineers with helpful information to figure out the issue during the function development process.
\end{itemize}

\subsubsection{User Interface}
QFaaS offers two main ways for quantum software engineers and end-users to interact with the core components using a command-line interface (QFaaS CLI) or a friendly user interface (QFaaS Dashboard).
    \begin{figure}[htbp]
        \centering
        \includegraphics[scale=0.33]{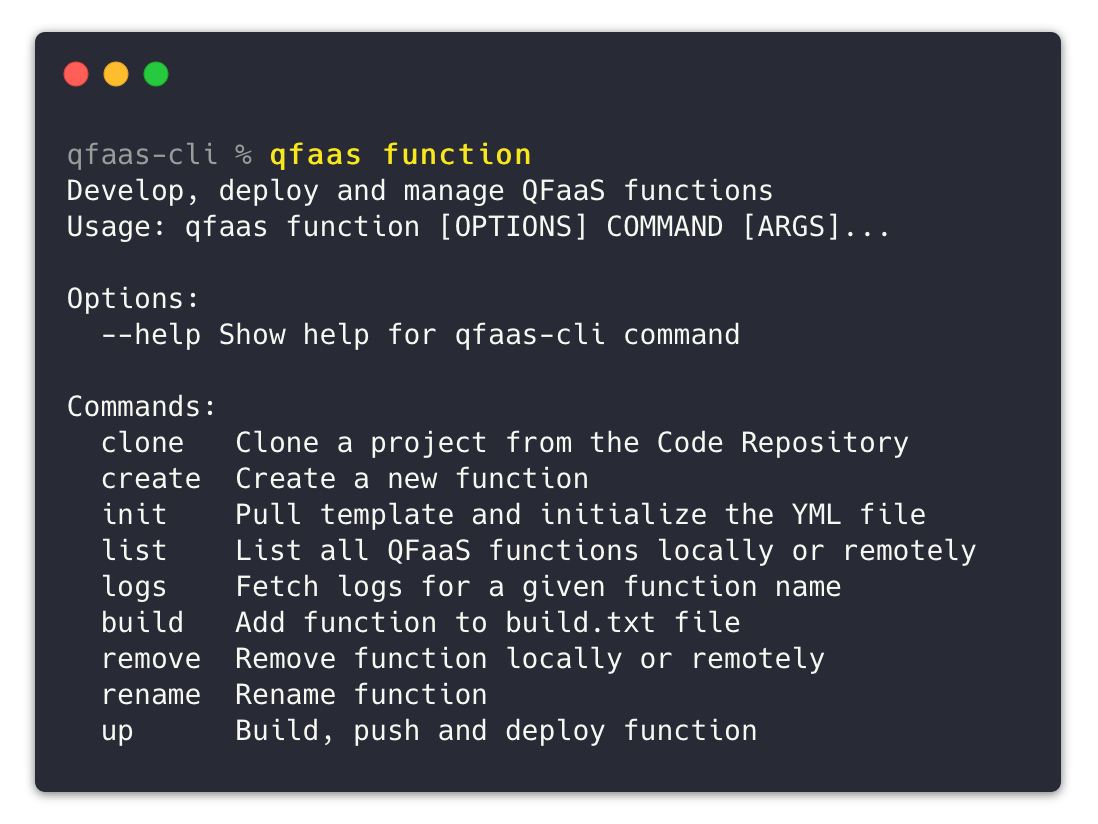}
        \caption{QFaaS CLI tool (sample commands for working with QFaaS functions)}
        \label{fig:qfass-cli}
    \end{figure}
\begin{itemize}
    \item \textbf{QFaaS CLI} tool (shown in Figure \ref{fig:qfass-cli}) is a Python-based command-line (CLI) tool for working with local function development environment. This CLI tool is mainly built for quantum software engineers to easily use their local IDE (Integrated Development Environment) such as Visual Code, Atom, and PyCharm to develop the functions and interact with the QFaaS core system remotely. The QFaaS CLI’s features are similar, corresponding with the supported features offered by the QFaaS Dashboard for quantum software engineers.
    \begin{figure}[htbp]
    \centering
    \includegraphics[scale=0.22]{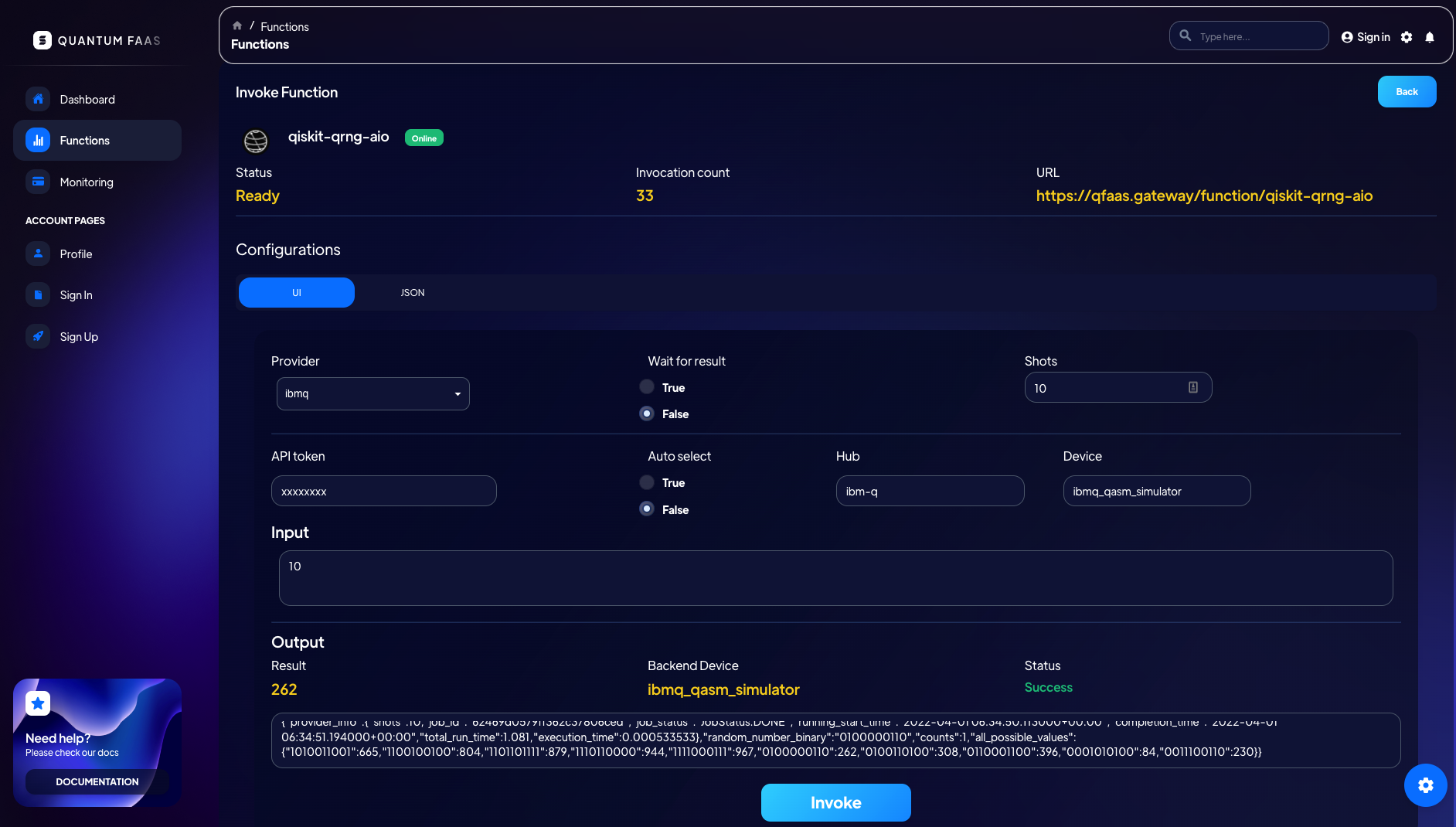}
    \caption{QFaaS Dashboard Example for invoking a Qiskit function with IBM Quantum provider}
    \label{fig:qfass-dashboard}
\end{figure}
    \item \textbf{QFaaS Dashboard} (shown in Figure \ref{fig:qfass-dashboard}) is a modern web-based user interface built by using ReactJS\footnote{https://reactjs.org/} for the frontend and Python 3.10 for its API backend (QFaaS Core APIs). This dashboard allows:
    \begin{enumerate}
        \item \textit{Quantum software engineers} to develop functions using a built-in code editor or upload their code files; update and manage their functions, templates; monitor the status of function deployment or overall system.
        \item \textit{End-users} to use the deployed quantum function by invoking (i.e., sending requests), monitoring, and managing their requests and results.
    \end{enumerate}

\end{itemize}

\subsection{Hybrid Quantum-Classical Function Life Cycle}
\label{qfaaslifecycle}
Inspired from several proposed Quantum Software Life Cycles \cite{Weder2020TheLifecycle, Weder2021QuantumLifecycle,Fu2021Quingo:Features}, we propose an altered 6-phase life cycle of a hybrid quantum-classical function (see Figure \ref{fig:qfass-fn}) for QFaaS as follows:

\begin{figure}[htbp]
    \centering
    \includegraphics[scale=0.097]{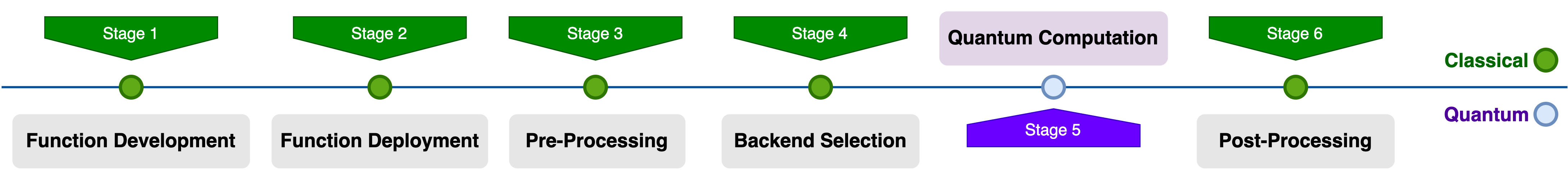}
    \caption{Hybrid Quantum-Classical Function Life Cycle.}
    \label{fig:qfass-fn}
\end{figure}

\begin{enumerate}
    \item \textbf{Function Development}: Quantum Software Engineers develop their quantum function, using quantum SDKs for quantum computation and optionally including classical parts for pre-processing, post-processing, and backend selection. After finishing the development, functions are pushed to a Git-based code repository and the function deployment process is triggered.
    \item \textbf{Function Deployment}: The quantum function is deployed at the classical backend (e.g., Kubernetes cluster) by employing modern software engineering technologies and models, such as containerization, continuous integration, and continuous deployment (CI/CD).
    During the deployment process, some components of the function could be compiled. For example, we use \textit{dotnet} to compile Q\# code into a classical binary, which could be imported into the main function handler. After deployment, the function is published as a service, allowing end-users access through the API gateway.
    \item \textbf{Classical Pre-Processing}: The function executes the classical computation task to pre-process the user’s input data and pass all the requested data to the appropriate component in the function handler.
    \item \textbf{Backend Selection}: An appropriate quantum backend must be selected for the quantum execution based on user preference. These quantum backends could be a quantum simulator or a physical quantum computer provided by a quantum cloud computing vendor (such as IBM Quantum, Amazon Braket, and Azure Quantum). In the case of simulators, it could be either the built-in simulator of the corresponding quantum SDK, which runs on top of a classical computer at the Kubernetes cluster (internal quantum simulators), or an external quantum simulator provided by the quantum cloud provider (such as IBMQ QASM Simulator, Amazon Braket SV Simulator).
    \item \textbf{Quantum Computation}: A corresponding quantum circuit will be built and then sent to the selected quantum backend according to user input. Suppose the end-user does not specify a specific quantum backend, QFaaS could automatically choose the best suitable (least busy) backend at the quantum provider to send the quantum circuit for execution. As today's quantum computers are NISQ devices \cite{Preskill2018QuantumBeyond}, each quantum execution should be run many times (shots) to mitigate the quantum errors. Besides, due to the limited number of available quantum computers, a quantum task (job) needs to be queued at the quantum provider (from seconds to hours) before execution. After the quantum computation is finished, the outcome is sent back to the function handler for post-processing or directly returned to the end-users.
    \item \textbf{Classical Post-Processing}: This optional step takes place at classical computation nodes to process the outcome from the quantum backend before sending it to the end-user via the API gateway.
\end{enumerate}

\subsection{Operation Workflows of QFaaS}
\subsubsection{Workflow for developing and deploying quantum functions}
We simplify the function development process for quantum software engineers by using QFaaS. They can follow these workflows to create a new function, update an existing function or figure out the issues during the development process:
\begin{itemize}
    \item \textbf{Develop a new function}: Figure \ref{fig:qfass-dev} illustrates the function developing process, including seven main steps. The first two steps are the responsibilities of the quantum software engineer, and the QFaaS framework handles the remaining steps.
    Our simple 2-step procedure to develop a new quantum function is as follows:
    \begin{enumerate}
        \item Create a new function by using the QFaaS Dashboard or the CLI tool:
        \begin{itemize}
            \item Specify which quantum templates will be used (Qiskit, Cirq, Q\#, or Braket); environment variables (such as secrets, annotations, scale factor) for the function.
            \item Compose Quantum functions using local IDE or the web-based IDE of QFaaS Dashboard.
        \end{itemize}
        \item Push function codes to the Code Repository through the QFaaS API Gateway.
        
        \textit{After these first two steps, QFaaS will automatically take responsibility for the rest of the deployment procedure by performing the following steps:}
        \item The API gateway forwards the function code and pushes it to the Code Repository.
        \item This triggers the Automation components to start the deployment process. We implemented the QFaaS Automation component on top of Gitlab Runner\footnote{https://docs.gitlab.com/runner/}.
        \item Pull the function template and combine it with function code to build up and containerize it into a Docker image. Then, those images will be pushed to a Container Registry to be stored for further utilization (such as migrating or scaling up a function).
        \item Deploy the function into the Kubernetes cluster at the classical cloud layer.
        \item Expose the service API URL endpoint corresponding to the deployed function. After this stage, the function serves as a service and is ready for invoking from end-users.
    \end{enumerate}
    \begin{figure}[htbp]
    \centering
    \includegraphics[scale=0.49]{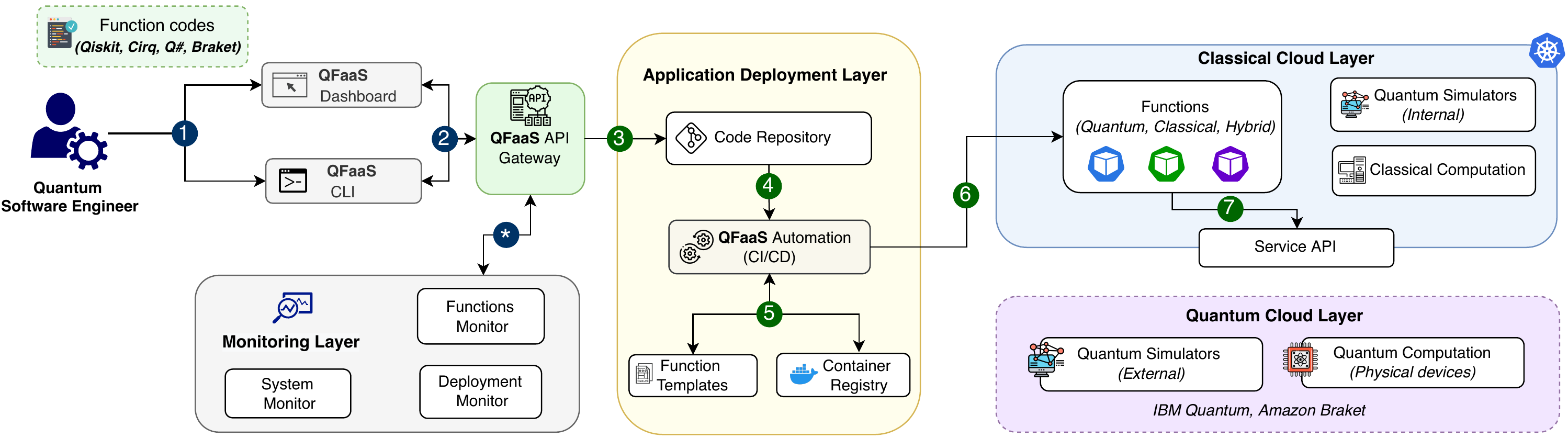}
    \caption{Operation workflow for developing and deploying QFaaS functions}
    \label{fig:qfass-dev}
\end{figure}
    \item \textbf{Update, delete functions}: QFaaS facilitate these actions for quantum software engineer by providing corresponding features at the Dashboard and CLI tool. After an update is triggered at the Function Development layer, QFaaS will activate the Automation component and take over the remaining tasks.
    \item \textbf{Monitor functions and system status}: The Monitoring layer periodically checks the status of the deployment process, function usage, and system status (such as network status, Kubernetes cluster resource consumption) to provide insights for quantum software engineers on demand.
    \item \textbf{Troubleshoot and diagnose the problem}: If any issues are discovered during the deployment or invocation, QFaaS will check and provide detailed logs for the engineer to investigate and figure out the problem. Once engineers fix all issues, it will automatically deploy to the cluster to ensure the continuous integration and continuous delivery of the latest version of the function for end-users.
\end{itemize}

\subsubsection{Workflow for invoking quantum functions}
The end-users could invoke (i.e., send their request to) the deployed function in many ways: 1) using the QFaaS Dashboard, 2) using the QFaaS CLI tool, or 3) integrating and calling QFaaS APIs from other applications or third-party API testing tools (such as Postman\footnote{https://www.postman.com/}).

\begin{figure}[htbp]
    \centering
    \includegraphics[scale=0.55]{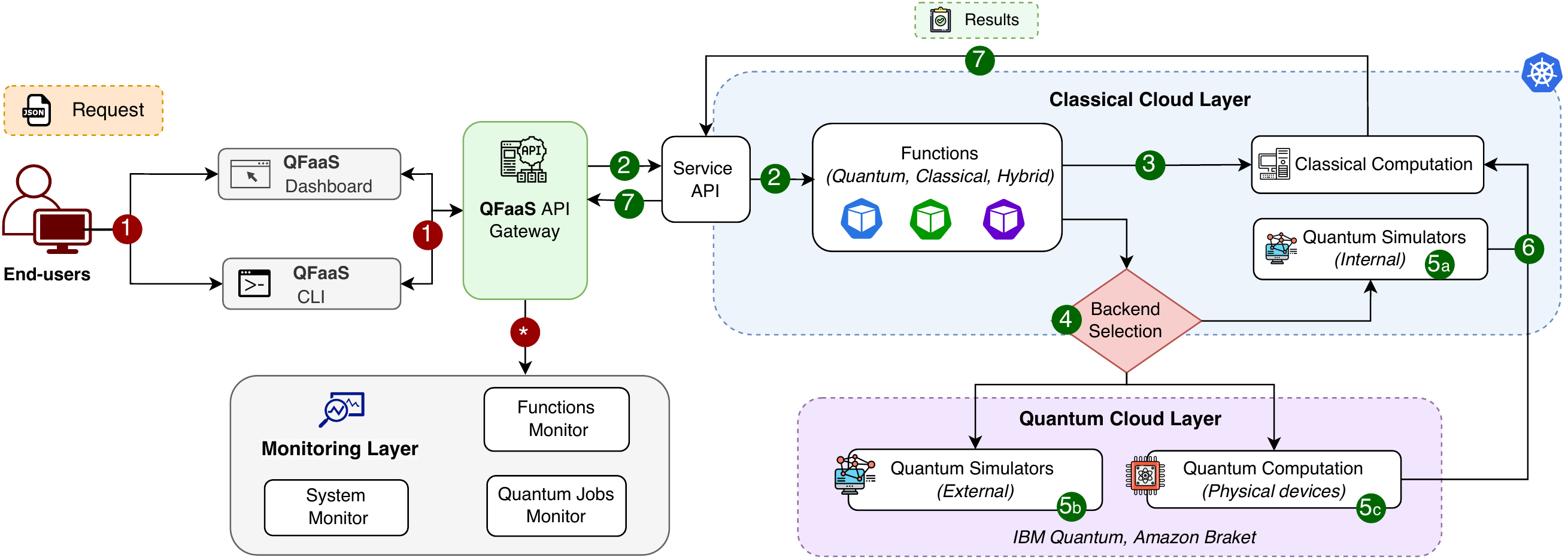}
    \caption{Operation workflow for invoking QFaaS functions}
    \label{fig:qfass-use}
\end{figure}

Figure \ref{fig:qfass-use} demonstrates the overall workflow for the function invocation, including seven main steps. In this procedure, end-users need to send their request (the first and only step) and then wait for the QFaaS framework to take responsibility for the remaining tasks.

\begin{enumerate}
    \item \textbf{Sending request}: In the requested data, user can clarify their preferred backend or let the framework automatically select the suitable backend, the result retrieval method, and the number of shots they want to repeat the quantum task. We describe the detailed sample request in section \ref{sec:fn-invocation}.

\textit{ After receiving the user’s requested data, QFaaS will automatically do the rest of the process.}
   
    \item \textbf{Routing the requests}: The API Gateway routes user requests to appropriate available functions. Suppose the function is not started yet or is being scaled down to zero. In that case, QFaaS will initialize and activate this function to process the user request (this situation is also known as \textit{cold-start} in serverless terminology).
    \item \textbf{Pre-processing} \textit{(optional)}: User’s input data is pre-processed at classical computation node.
    \item \textbf{Backend Selection}: An appropriate backend is selected based on user requests and the availability at the quantum provider.
    \item \textbf{Executing the quantum job}: A corresponding user’s input data is generated and sent to the selected backend device. The quantum backend device could be an internal quantum simulator (5a), an external quantum simulator (5b), or a physical quantum computer (5c). Then, that circuit is compiled to the appropriate quantum system language (such as QASM), and the quantum backend performs the quantum task. After finishing the execution, the outcome from the quantum backend will be sent back to the function handler for further processing. Suppose end-users want to track the response data later on, the function will send back the quantum Job ID and information of the backend device after successfully submitting the quantum circuit to the quantum backend.
    \item \textbf{Post-processing} \textit{(optional)}: The outcome from quantum backends could be analyzed and post-processed before sending back to end-users and storing it to the database.
    \item \textbf{Returning the results}: After the previous step, the final result will be returned to end-users via the API Gateway, as the same way when they submit the request. 
    End-users can get the result data, and information about the backend device is used for the quantum execution. The quantum engineers could also add other information and further analysis to help end-users get insight from the execution result.
\end{enumerate}

\section{Design and Implementation of QFaaS}
\subsection{QFaaS Core APIs}
\label{section-coreapi}
This section provides a detailed design and implementation of the QFaaS Core APIs, which take responsibility for primary functionalities in the QFaaS framework. We have developed this API set using Python 3.10 with FastAPI\footnote{https://fastapi.tiangolo.com/}, a high-performance Python-based framework supporting the Asynchronous Server Gateway Interface (ASGI) for concurrent execution. We also used the MongoDB\footnote{https://www.mongodb.com/} database to store the persistent data as JSON schemes.

\begin{figure}[htbp]
    \centering
    \includegraphics[scale=0.12]{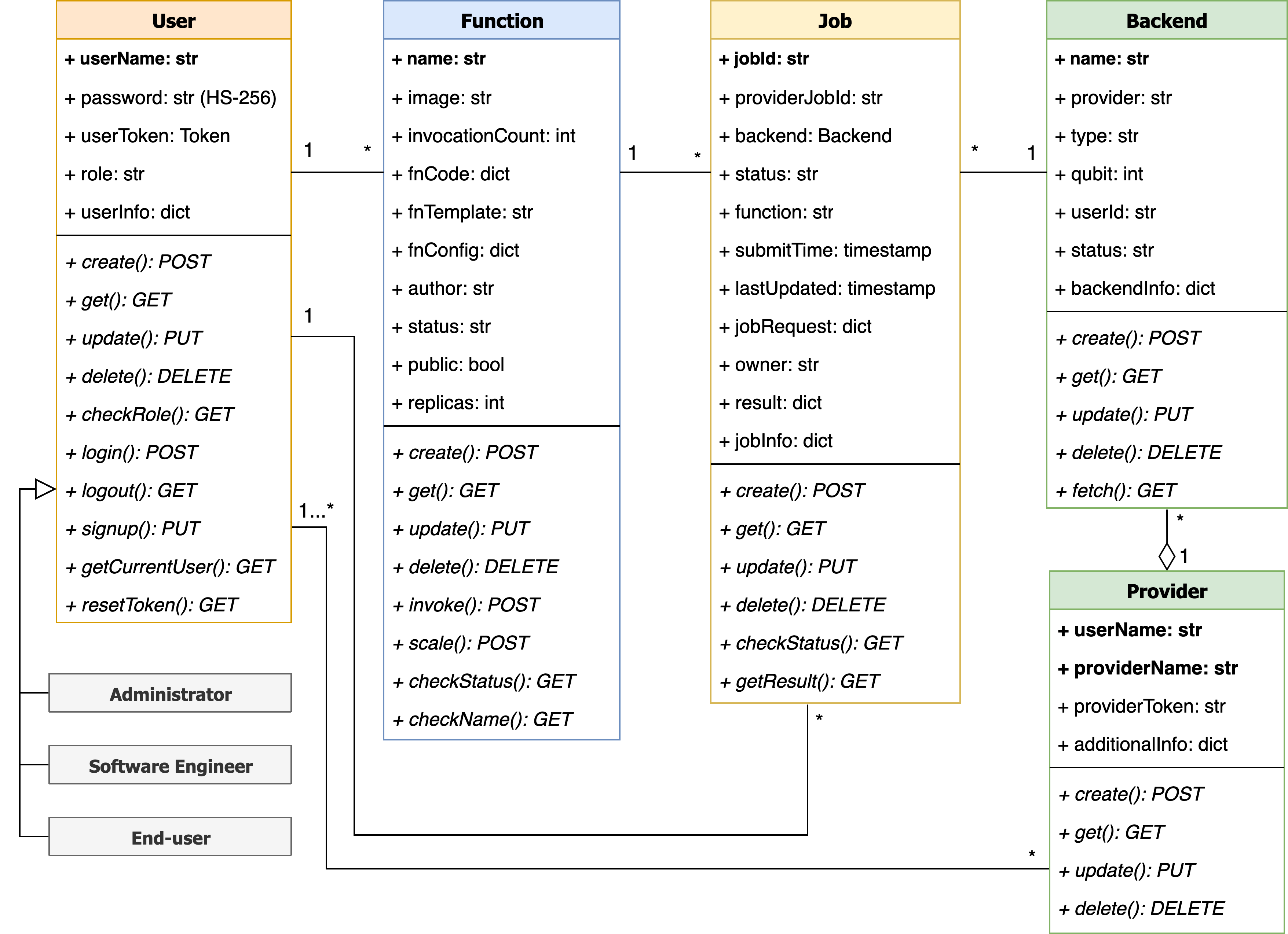}
    \caption{Overall Class Diagram of QFaaS Core REST APIs}
    \label{fig:qfass-coreapi}
\end{figure}

Figure \ref{fig:qfass-coreapi} depicts the overall class diagram, with attributes and methods of each object in QFaaS Core APIs. 

All objects are associated with four essential CRUD operations: Create - Read (get) - Update and Delete, using the proper HTTP methods, namely POST, GET, PUT, and DELETE, respectively.
\begin{itemize}
    \item \textbf{User:} This class defines user attributes and methods to facilitate access control features. We categorized three different users: administrator, software engineer, and end-user with different privileges in the system. Administrators have complete control of all components; software engineers can develop and deploy functions, while end-users can only use their appropriate functions. Each active user is assigned a unique token (using OAuth2 Bearer specification\footnote{https://datatracker.ietf.org/doc/html/rfc6750}), which is used for authentication, authorization, and dependency check for each interaction with the core components of the QFaaS, such as creating a new function or invoking a function. This implementation enhances security for the whole framework and provides a multiple-user environment for taking advantage of the QFaaS framework.
    \item \textbf{Function:} The Function class defines each function's properties and supported methods. Each function belongs to a specific software engineer (author) and could be published for the end-user. The \textit{CRUD}, \textit{invoke()} and \textit{scale()} methods of this object interacts directly with other architectural components such as Code Repository, Container Registry and Kubernetes Cluster to handle the function deployment, management, and invocation.
    \item \textbf{Job:} A job in QFaaS is a computation task submitted to a quantum backend (the internal cluster or external providers) for execution. All properties and methods of a job are defined in the Job class. Each Job could have two unique IDs: \textit{jobID} assigned by QFaaS associated with a \textit{providerJobID} given by an external provider for further job monitoring. The function invocation initializes the job object. After finishing the execution, job results are stored in the NoSQL database for further investigation or post-processing.
    \item \textbf{Provider:} The provider class handles a user's authorization to external quantum providers (such as IBM Quantum and Amazon Braket). The design of this class ensures that each user has the specific privilege to access their quantum providers only.
    \item \textbf{Backend:} A backend is a device, such as a classical computer, a quantum simulator, or a quantum computer, which takes responsibility for the job execution. The Backend class defines the attributes and methods to interact with the backend provided by the classical cluster or external quantum providers.
\end{itemize}

\subsection{QFaaS Function Development}
\subsubsection{Function Templates}
Quantum function templates are Docker images containing the environmental setup for quantum function development. Each quantum SDK has a corresponding \code{Dockerfile}; which specifies all necessary libraries and packages for that SDK. These templates inherit the \textit{of-watchdog}\footnote{https://github.com/openfaas/of-watchdog} component for initiating and monitoring functions. In this way, quantum functions could be kept \textit{“warm”} (i.e., running) with low latency and maintain persistent HTTP connections, quickly serving the user’s request.

\subsubsection{Function Structure}
Each quantum function has a simple working directory, including main components following the common pattern of the serverless platform (such as AWS Lambda \cite{awslambda}):
\begin{itemize}
    \item \textbf{Function dependencies:} We can declare all necessary Python-based libraries in the \code{requirements.txt} file. These libraries will be automatically installed during the function deployment and could be imported for use within the function.
    \item \textbf{Function handler}: include the source code for the function, including classical parts (using Python) and quantum parts (using Qiskit, Cirq, Q\#, and Braket). When end-users invoke the function, QFaaS executes the function handler and starts the computation at an appropriate backend device. Handler for Qiskit and Cirq function could be defined at \code{handler.py} file while Q\# function requires us to define an additional Q\# code at \code{handler.qs} file and then import it to the main \code{handler.py} file. The sample format for each function handler to handle user requests and return the response, with classical pre-processing and post-processing parts, is defined using the general Python syntax as Code \ref{lst:fn-code}.
\end{itemize}
First, we import all necessary quantum libraries (and Q\# operation for Q\# function only). Then, we define the function methods as follows:
\begin{itemize}
    \item \textbf{Classical pre-processing and post-processing}: In each function, we can optionally define the pre-processing and post-processing methods executed on classical computers (Kubernetes cluster).
    
    \item \textbf{Backend selection}: We can also define the custom strategy to allow end-users to select the appropriate quantum backend (internal simulator, external simulator, or quantum computer at cloud providers).
    \item \textbf{Main handler method}: The primary handle method is similar to AWS Lambda \cite{awslambda}, while the user’s requests are delivered to the function in two objects: \code{event} and \code{context}. The \code{event} is JSON-based data that comprise user input data, followed by the QFaaS sample format, while \code{context} provides HTTP methods (such as GET and POST), HTTP headers, and other properties, which are optional for the request. After finishing all the processing, the function handler could return the result to end-users, including the HTTP Status Code and response data in JSON format. QFaaS uses JSON as a default format to standardize user requests and responses, allowing the quantum software engineer to customize their format if needed.
\end{itemize}
\begin{lstlisting}[language=Python, caption=Sample structure of a hybrid quantum-classical function, label={lst:fn-code}]
import qfaas
import <additional_libraries>

def pre_processing(data):
  [pre processing method]

def post_processing(data):
  [post processing method]

def backend_selection(input):
  [quantum backend selection method]

def handle(event, context):
  [main function handler]
  return result
\end{lstlisting}

\subsection{QFaaS Core Functionalities Implementation}
In this section, we describe the implementation of two core functionalities in the QFaaS framework, including function deployment and function invocation.
\subsubsection{Function Deployment}
The overall function deployment procedure is implemented following Algorithm \ref{algo:fn-deployment}. 

\begin{figure}[htbp]
    \centering
    \includegraphics[scale=0.24]{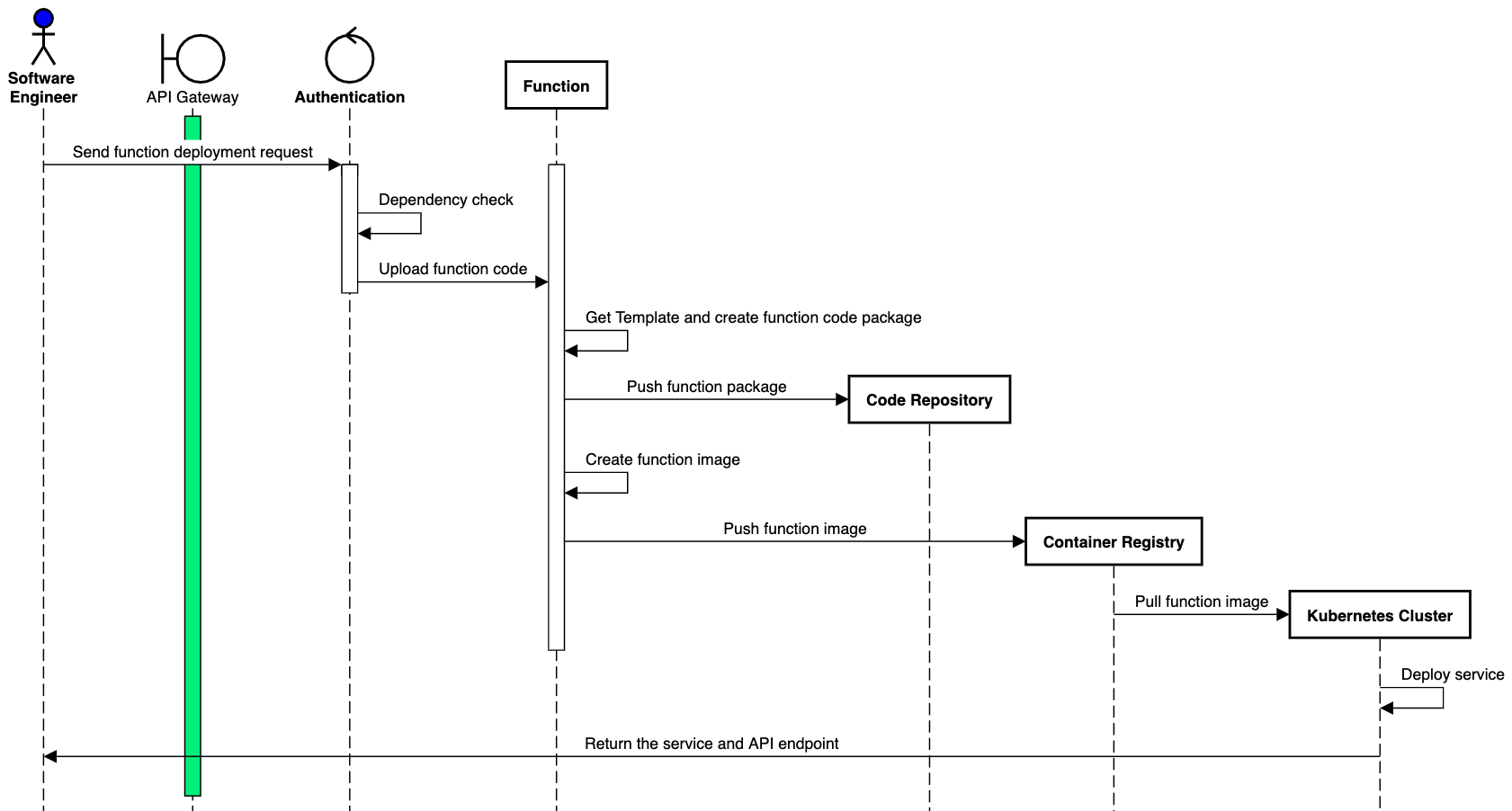}
    \caption{Sequence Diagram of function deployment process in QFaaS}
    \label{fig:qfass-seqdep}
\end{figure}

Before launching the deployment process, the \code{dependencyCheck()} validates the access token and checks the permission of the current user, who is sending the function deployment request. The deployment process is only triggered if the \code{dependencyCheck()} is passed and the function name is valid. First, the function code template is retrieved based on the user selection. Then, the function codes are integrated into that template to create a function package and pushed to the Code Repository for versioning control purposes.
The Automation component will be triggered to perform the \code{ContinuousDeployment} process whenever new function codes are updated (lines 8 - 13). This process starts by containerizing the function package into an appropriate Docker image. Then, it uploads that image to the Container Registry to facilitate the continuous deployment process. Afterward, a container-based service is deployed into the Kubernetes cluster, and a corresponding API endpoint is published. Finally, this service is ready for invocation from authorized users through the QFaaS API gateway. The sequence diagram of a sample function deployment process is shown in Figure \ref{fig:qfass-seqdep}.
\begin{algorithm}
  \small
    \SetKwData{fnCode}{fnCode}
    \SetKwData{fnName}{fnName}
    \SetKwData{fnImage}{fnImage}
    \SetKwData{function}{function}
    \SetKwData{template}{template}
    \SetKwData{fnTemplate}{fnTemplate}
    \SetKwData{fnConfig}{fnConfig}
    \SetKwData{fnService}{fnService}
    \SetKwData{fnEndpoint}{fnEndpoint}
    \SetKwData{currentUser}{currentUser}
    
    \SetKwFunction{getCurrentUser}{getCurrentUser}
    \SetKwFunction{dependencyCheck}{dependencyCheck}
    \SetKwFunction{fnNameCheck}{fnNameCheck}
    \SetKwFunction{ContinuousDeployment}{ContinuousDeployment}
    \SetKwFunction{pushImageToContainerRegistry}{pushImageToContainerRegistry}
    \SetKwFunction{pushFnPackageToCodeRepository}{pushFnPackageToCodeRepository}
    \SetKwFunction{getTemplate}{getTemplate}
    \SetKwFunction{createFunctionPackage}{createFunctionPackage}
    \SetKwFunction{buildFunctionImage}{buildFunctionImage}
    \SetKwFunction{deployServiceToCluster}{deployServiceToCluster}
    \SetKwFunction{publishServiceAPI}{publishServiceAPI}
        
    \SetKwInOut{KwIn}{Input}
    \SetKwInOut{KwOut}{Output}
    
    \SetKwProg{procedure}{procedure}{}{}

    \KwIn{\fnName: function name, \fnCode: main handler code, \fnTemplate: template name, \fnConfig: additional configuration}
    \KwOut{\fnService: Deployed service of given function, \fnEndpoint: API endpoint URL of deployed function}

    \currentUser $\leftarrow$ \getCurrentUser() \\
    \eIf{\dependencyCheck(\currentUser) is passed}{
        \eIf{\fnNameCheck(\fnName) is valid}{
            \template $\leftarrow$ \getTemplate(\fnTemplate) \\
            \function $\leftarrow$ \createFunctionPackage(\fnCode, \template, \fnConfig) \\
            \pushFnPackageToCodeRepository(\function) \\
            \textit{Trigger the Continuous Deployment} \\ 
            \procedure{\ContinuousDeployment}{
                \fnImage $\leftarrow$ \buildFunctionImage(\function) \\
                \pushImageToContainerRegistry(\fnImage) \\
                \fnService $\leftarrow$ \deployServiceToCluster(\fnImage) \\
                \fnEndpoint $\leftarrow$ \publishServiceAPI(\fnService)
            }{\KwRet{\fnService, \fnEndpoint}}
        }{\KwRet{$FunctionNameError$}}
    }{\KwRet{$PermissionError$}}
    \caption{QFaaS Function Deployment}
    \label{algo:fn-deployment}
\end{algorithm}

\subsubsection{Function Invocation}
\label{sec:fn-invocation}
We describe the implementation of handling the function invocation in Algorithm \ref{algo:fn-invocation}. 

Similar to the function deployment, we perform the \code{dependencyCheck()} as a mandatory requirement before each invocation to ensure the framework's security. After passing that validation, the QFaaS API gateway forwards the request to the corresponding service in the cluster. 
\begin{algorithm}
    \small
    \SetKwData{req}{req} \SetKwData{fnEndpoint}{fnEndpoint}
    \SetKwData{result}{result} \SetKwData{jobID}{jobID}
    \SetKwData{currentUser}{currentUser}
    \SetKwData{function}{function}
    \SetKwData{processedData}{processedData}
    \SetKwInOut{KwIn}{Input}
    \SetKwInOut{KwOut}{Output}
    \SetKwData{qcircuit}{qcircuit}
    \SetKwData{pToken}{pToken}
    \SetKwData{backend}{backend}
    \SetKwProg{procedure}{procedure}{}{}
    
    \SetKwFunction{forwardRequest}{forwardRequest}
    \SetKwFunction{preProcessing}{preProcessing}
    \SetKwFunction{buildQuantumCircuit}{buildQuantumCircuit}
    \SetKwFunction{getProviderToken}{getProviderToken}
    \SetKwFunction{BackendSelection}{BackendSelection}
    \SetKwFunction{getFunction}{getFunction}
    \SetKwFunction{submitJob}{submitJob}
    \SetKwFunction{postProcessing}{postProcessing}
    \SetKwFunction{getID}{getID}

    \KwIn{\req: User request data (JSON), \fnEndpoint: API endpoint of function}
    \KwOut{\result: Computation result data, or \jobID: Job ID for later tracking }

    \currentUser $\leftarrow$ \getCurrentUser() \\
    \eIf{\dependencyCheck(\currentUser) is passed}{
        \function $\leftarrow$ \getFunction(\fnEndpoint) \\
        \forwardRequest(\req, \function) \\
        \processedData $\leftarrow$ \preProcessing(\req.data) \\
        \qcircuit $\leftarrow$ \buildQuantumCircuit(\processedData) \\
        \pToken $\leftarrow$ \getProviderToken(\currentUser) \\
        \backend $\leftarrow$ \BackendSelection(\qcircuit.qubit, \req.backendInfo, \pToken) \\
            job $\leftarrow$ \submitJob(\currentUser, \qcircuit, \backend, \req.shots) \\
            \eIf{\req.waitForResult == \textsf{True}}{
                \result $\leftarrow$ \postProcessing(job.result) \\
                \KwRet{\result}
            }{
                \jobID $\leftarrow$ \getID(job) \\
                \KwRet{\jobID}
            }
        
    }{\KwRet{$PermissionError$}} 
    \caption{QFaaS Function Invocation}
    \label{algo:fn-invocation}
\end{algorithm}

We define a sample JSON format for the request and response format in the function template. However, quantum software engineers can customize the request and response to adapt to their functional requirements. A sample JSON request is as follows (Code \ref{lst:invoke-rqst}):
    \begin{lstlisting}[language=json, caption=Sample JSON request format to invoke a QFaaS function, label={lst:invoke-rqst}]
{
    "input": <input data>,
    "provider": <provider name>,
    "shots": <number of shots>,
    "waitForResult": <true or false>,
    "backendInfo": {
		    "autoselect": <true or false>,
		    "type": <expected backend name if autoselect = true>,
		    "backendName": <quantum backend name if autoselect = false> }
}\end{lstlisting}
    Apart from the input data, end-users can also define:
    \begin{itemize}
        \item \code{provider}: clarify their preferred quantum backend, either an internal simulator or quantum cloud provider (including IBM Quantum and Amazon Braket). In the case of selecting an external quantum cloud provider, they need to specify the information of their preferred quantum backend in the \code{backendInfo} section. 
        If the \code{autoselect} variable is \code{true} and the expected backend \code{type} (internal/external simulator or quantum computer) is set, QFaaS will automatically select the best-suited quantum backend (least busy devices with enough qubits for the computation). The users can also manually select a specific quantum backend by setting this value to \code{false} and declaring the backend name in the \code{backendName}.
        \item \code{shots}: number of repetition times they want the quantum computation to perform
        \item \code{waitForResult}: if this option is set to true, they need to wait at the current session until the function execution is done and receive the result. This waiting time at the quantum cloud provider could be very long (for queuing, transpiling, and executing) due to the current NISQ nature. Otherwise, they have another option by letting QFaaS send their job to the quantum backend device and providing them the Job ID to track the result later. 
    \end{itemize}

Based on the functionalities of that service, the pre-processing and post-processing could be performed before and after the execution. Following that optional data processing, a corresponding quantum circuit is built based on the user’s input data. Then, the \textit{BackendSection} will be performed as Algorithm \ref{algo:backendselection}.

\begin{algorithm}
    \small
    \SetKwData{rQ}{rQ}
    \SetKwData{be}{be}
    \SetKwData{token}{token}
    \SetKwData{backend}{backend}
    \SetKwInOut{KwIn}{Input}
    \SetKwInOut{KwOut}{Output}
    \SetKwData{beList}{beList}
    \SetKwData{provider}{provider}
    
    \SetKwFunction{getInternalBackend}{getInternalBackend}
    \SetKwFunction{getProvider}{getProvider}
    \SetKwFunction{getBackend}{getBackend}
    \SetKwFunction{getLeastBusyBackend}{getLeastBusyBackend}
    \SetKwFunction{getBackendList}{getBackendList}
    
    \SetKwProg{procedure}{procedure}{}{}

    \KwIn{\rQ: number of required qubit, \be: backend info, \token: Token to access external provider (optional)}
    \KwOut{\backend: quantum backend object }
    
    \procedure{BackendSelection (\rQ, \be, \token):}{
        \backend $\leftarrow$ null \\
        \eIf{\be.internal == true}{
            \backend $\leftarrow$ \getInternalBackend(\be.backendName)
        }{
            \provider $\leftarrow$ \getProvider(\token) \\
            \beList $\leftarrow$ [] \\
            \For{b in \provider.\getBackendList())}{
              \If{b.qubit $\ge$ \rQ and b.operational == \textsf{True} and b.type in \be.type}{
                        \beList $\xleftarrow{+}$ b
                    }  
            }
            \eIf{\be.autoselect == \textsf{True}}{
                \backend $\leftarrow$ \getLeastBusyBackend(\beList)
            }{
                \If{\be.backendName in \beList}{
                   \backend $\leftarrow$ \getBackend($\be.backendName$)
                }
            }
            }

    }{\KwRet{\backend}}
    \caption{QFaaS Backend Selection}
    \label{algo:backendselection}
\end{algorithm}

\begin{figure}[htbp]
    \centering
    \includegraphics[scale=0.27]{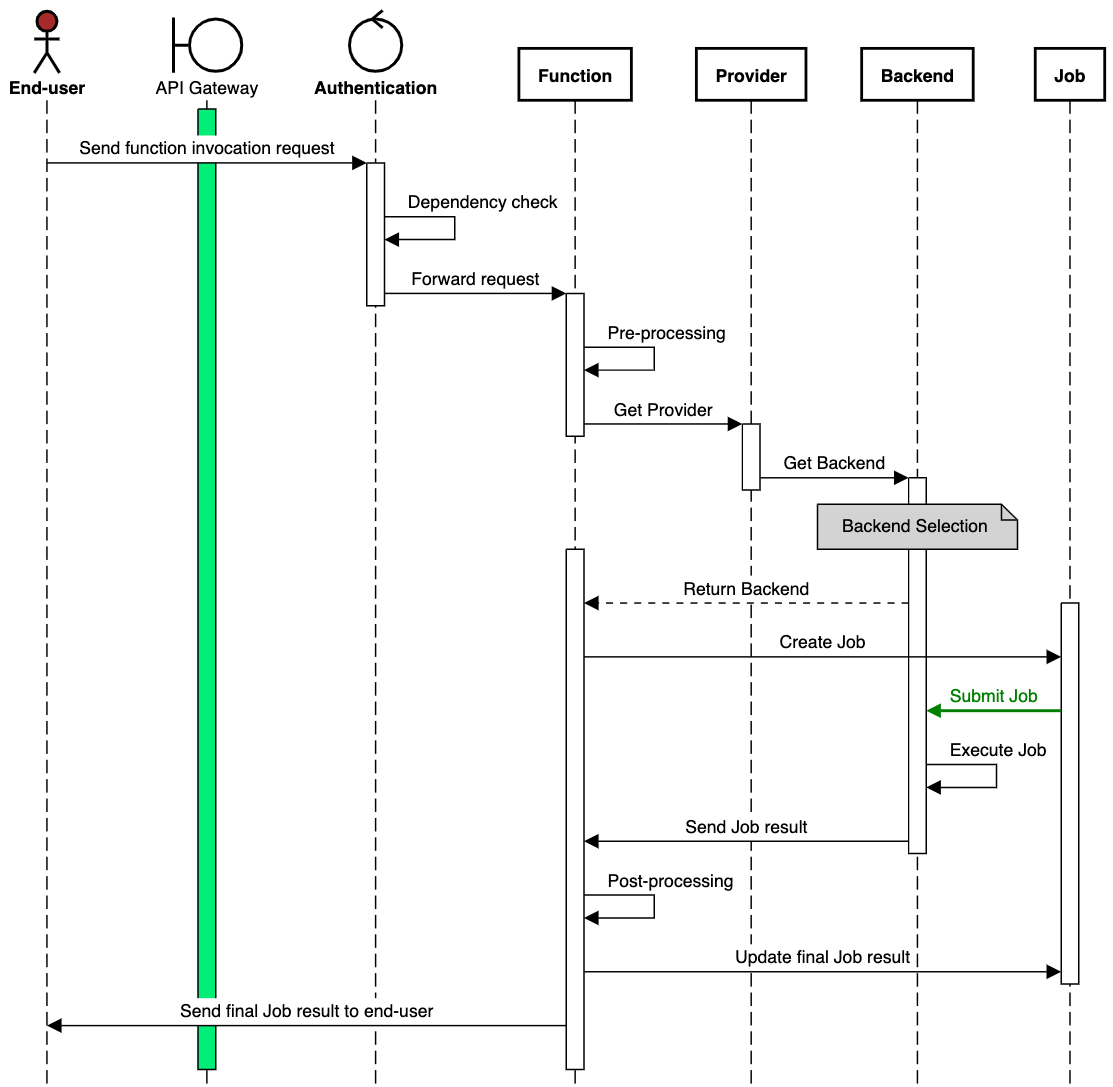}
    \caption{Sequence Diagram of function invocation process in QFaaS}
    \label{fig:qfass-seqinvoke}
\end{figure}

Given the number of required qubits for the quantum circuit, backend preference (such as backend type, manually or automatically selected), and provider access information, an appropriate backend object is returned. Users can allow the framework to automatically select the best-suited backend in a specific type, such as a quantum simulator, quantum computer, or any kind. In that case, QFaaS will inspect and select the least busy backend (i.e., the backend that has the shortest waiting queue) from the provider.
After a suitable backend is determined, the function invocation is continued by submitting the quantum circuit to that backend for execution. Then, based on user preference of whether or not to wait until receiving the result, QFaaS will either return the final result (after performing the post-processing - if any) or the job ID for further tracking of the result. The overall interaction between different objects during the function invocation process is shown in Figure \ref{fig:qfass-seqinvoke}.

\section{Performance Evaluation}
\subsection{Environment setup}
To validate the proposed framework, we deployed the core components of QFaaS on a set of four virtual machines (VMs) offered by the Melbourne Research Cloud\footnote{https://dashboard.cloud.unimelb.edu.au/}. We set up the Kubernetes cluster with Docker as underlying container technology on three VMs (one master node with 4 vCPU, 16GB RAM, and two worker nodes with 8vCPU, 32 GB RAM each). The QFaaS Code Repository and Automation components are deployed to the last VM (4 vCPU, 16 GB RAM). For the computation layer, we have tested the Qiskit functions with the QASM simulator on both the classical computers at the Kubernetes cluster and quantum backends at the IBM Quantum provider \cite{ibmq} (from the IBMQ hub at the University of Melbourne). For Braket functions, we used their local simulator and Amazon Braket quantum computing service (through the support of Strangeworks Backstage Pass program \cite{Strangeworks2022StrangeworksPlatform}). For Q\# and Cirq functions, we used their built-in quantum simulator and executed it on classical computers in the Kubernetes Cluster.

\subsection{Case study 1: Quantum Random Number Generators (QRNG)}
Random numbers play an essential role in cryptography, cybersecurity, finances, and many other scientific fields. By leveraging quantum principles, Quantum Random Number Generator (QRNG) has been proposed as a reliable way to provide truly randomness, which can not be achieved by using classical computers. This topic has been gaining much interest in the last 20 years \cite{Huang2021QuantumPlatform}. To give an example of how to generate quantum random numbers by utilizing the superposition state and demonstrate the workflow of QFaaS in action, we deployed a simple QRNG function in four different quantum SDKs and languages, including Qiskit, Cirq, Q\#, and Braket.

\subsubsection{Developing QRNG function in multiple quantum SDKs}
\begin{figure}[htbp]
    \centering
    \includegraphics[scale=0.48]{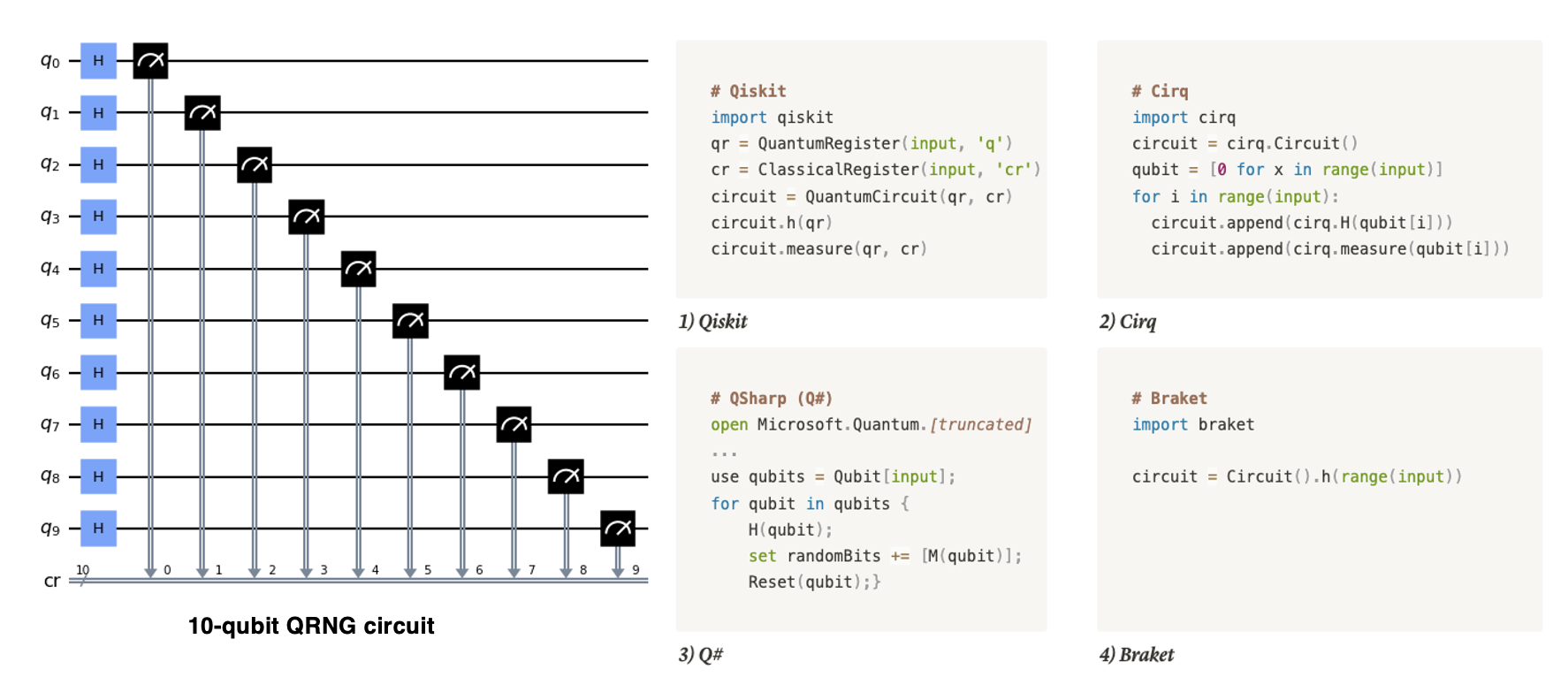}
    \caption{QRNG circuit for generating 10-bit random number (left - generated using Qiskit) and Sample code snippets to generate the quantum circuit with given integer \code{input} (right)}
    \label{fig:qfass-qrng}
\end{figure}

We used a simple quantum circuit for generating a random number with the number of qubits as the user input. The main idea of this circuit is to leverage the Hadamard gate to create the superposition state of each qubit and then measure to get a random value (0 or 1) with the same possibility (50\%). According to the user’s request, our functions will dynamically generate an appropriate quantum circuit with the corresponding qubits. Figure \ref{fig:qfass-qrng} shows an example of the quantum circuit for generating a 10-bit random number.


Since Qiskit, Cirq and Braket are Python-based SDKs, we need to write quantum code to develop a quantum function in the default \code{handler.py} file. When using Q\#, as it is an independent quantum programming language, we need to develop the function in a \code{handler.qs} file, and the .NET framework will compile this file. Fortunately, it also supports integrating Q\# with Python code, and we need to import the operation from the Q\# file to the \code{handler.py} file. QFaaS will take responsibility for the remaining procedure. 
We can also include several classical processing parts in each quantum function, especially for pre-processing the user input data and post-processing the result from the quantum computation layer. To validate this feature, we implemented simple post-processing for the QRNG function by analyzing all possible outcomes when the function is executed multiple times (shots) and returned the most frequent result to the end-user. The classical processing will be performed on the Kubernetes cluster, which runs on a classical computer after the quantum computation layer completes its processing.


\subsubsection{Deploying and invoking QRNG Functions}
After developing a quantum function, we upload it to QFaaS Code Repository using the QFaaS Dashboard or QFaaS CLI tool. Whenever the QFaaS Automation detects new updates from the Code Repository, it will automatically check, build the Docker images, push that image to the Docker registry and deploy it to the Kubernetes cluster. Before the deployment, we can also integrate some intermediate processes to verify source code quality or perform a security check.
Subsequently, QFaaS will release each quantum function with a unique URL accessible to end-users via the API gateway. User request data will be \code{jsonify} (i.e., converted to JSON format) and then sent to the API gateway. 




The request JSON for invoking the QRNG function using all supported SDKs and languages is similar. After finishing processing, the sample response is as Code \ref{lst:invoke-16b-rep}. This result indicates that the generated random number is 6493 (0001100101011101 in binary), one of the most frequent (2 times occurrence) random numbers generated by the Qiskit QRNG function. Thanks to the post-process, we also have other possible results after running this function 1024 times, i.e., 17990 and 26321.

\begin{lstlisting}[language=json, caption=Sample response data for returning a random 16-bit number from Qiskit QRNG function, label={lst:invoke-16b-rep}]
{
    "result": 6493,
    "backend_device": "ibmq_qasm_simulator",
    "detail": {
        "provider_info": {
            "shots": 1024,
            "job_id": "62301d63e6b7bb485520xxxx",
            "job_status": "DONE",
            "running_start_time": "2022-03-15 05:00:24.072000+00:00",
            "completion_time": "2022-03-15 05:00:25.093000+00:00",
            "total_run_time": 1.021},
        "random_number_binary": "0001100101011101",
        "counts": 2,
        "all_possible_values": {
            "0001100101011101": 6493,
            "0100011001000110": 17990,
            "0110011011010001": 26321 }
    }
}\end{lstlisting}

\subsubsection{Performance Evaluation on Quantum Simulators}
We conducted a series of experiments using the JMeter tool\footnote{https://jmeter.apache.org/} to benchmark the performance of the QRNG function in three different quantum simulators on the QFaaS framework. For a practically fair comparison, we used the default quantum simulator (QASM simulator for Qiskit, \textit{braket\_sv} simulator for Braket, and built-in simulator for Q\# and Cirq) of all frameworks for execution. We repeat each experiment 100 times, then measure the average response time and the standard deviation when executing the QRNG function using 1 qubit to 20 qubits in each quantum SDK.
\begin{figure}[htbp]
    \centering
    \includegraphics[scale=0.14]{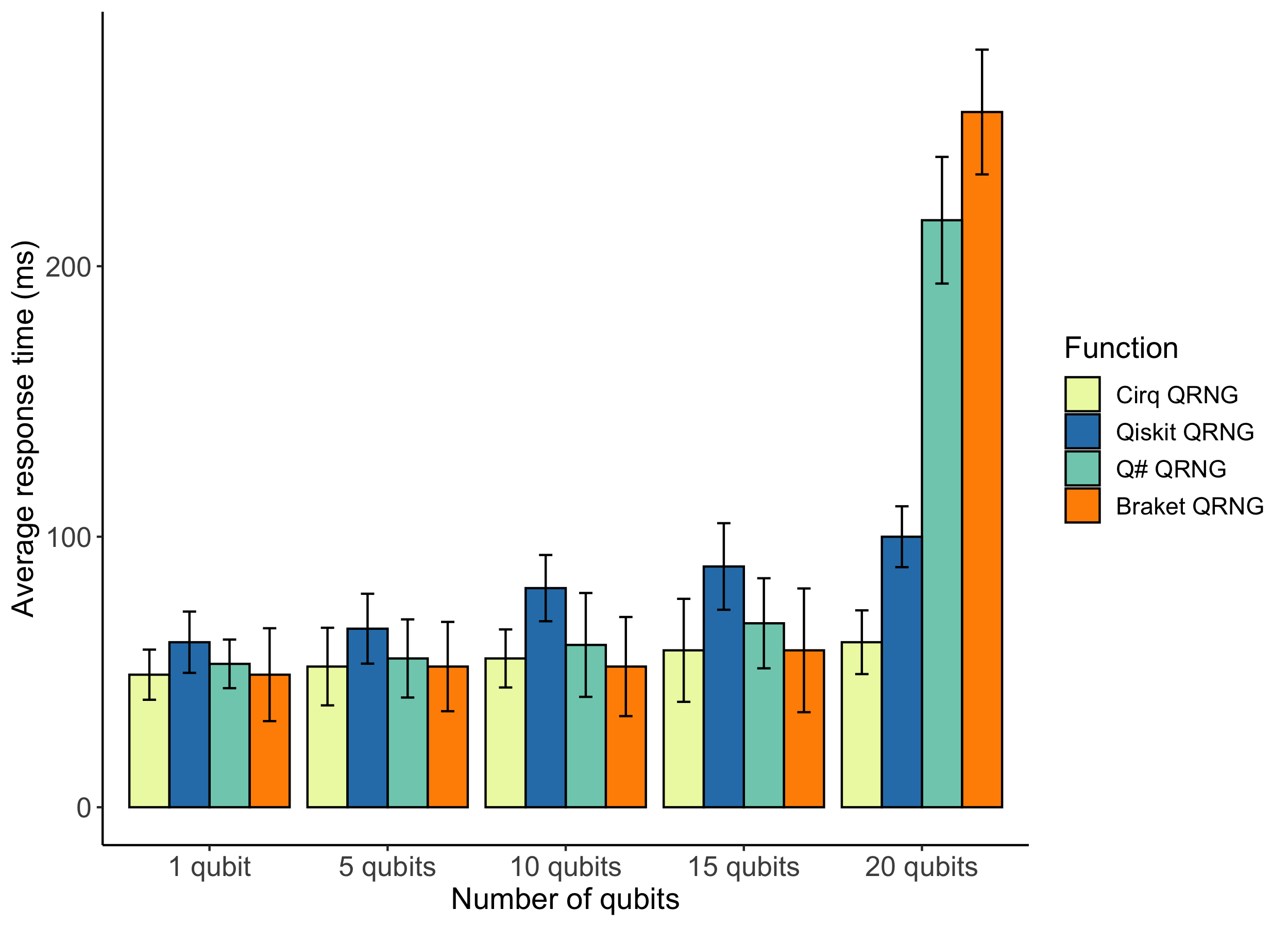}
    \caption{Average response time evaluation of QRNG function using the simulator of popular quantum SDKs and languages}
    \label{fig:qrng-eval}
\end{figure}

Figure \ref{fig:qrng-eval} illustrates the average response time of three functions when we increase the number of qubits from 1 to 20. The Cirq simulator registers the fastest response time in all test cases with a slight increase from 49 ms for generating a 1-qubit random number to 61 ms for a 20-qubit one. A similar trend could also be seen if we look at the Qiskit, Q\#, and Braket function figures when the number of qubit increases from 1 to 15. The Qiskit function response time is slightly longer than Cirq and Q\# during the 1- to 15-qubit period. However, when the number of qubits reaches 20, the response time of the Q\# and Braket functions increases significantly and doubles the Qiskit counterpart. This evaluation demonstrates the current state of several popular quantum simulators, but it depends on specific quantum applications and could be changed with the further development of these SDKs.

\subsubsection{Scalability evaluation}
We could enable the auto-scaling feature to scale up the deployment horizontally (i.e., increase the number of pods), dealing with the scenario when the request workload grows significantly.

To validate the scalability of our framework, we perform a set of evaluations on the 10-qubit Qiskit QRNG function. In this evaluation, we increase N - the number of concurrent users from 8 to 64, using the JMeter benchmarking tool. In each case, we conduct a set of three different scenarios: non-scale (1 pod/function), scale up to N/2 pods and scale up to N pods (we fixed the number of pods for evaluation purposes only). For example, suppose there are 64 users (N) invoking the function simultaneously. In that case, we will conduct three test cases: 1 pod, 32 pods (N/2), and 64 pods (N) and record the average response time and the standard deviation. 

\begin{figure}[htbp]
    \centering
    \includegraphics[scale=0.13]{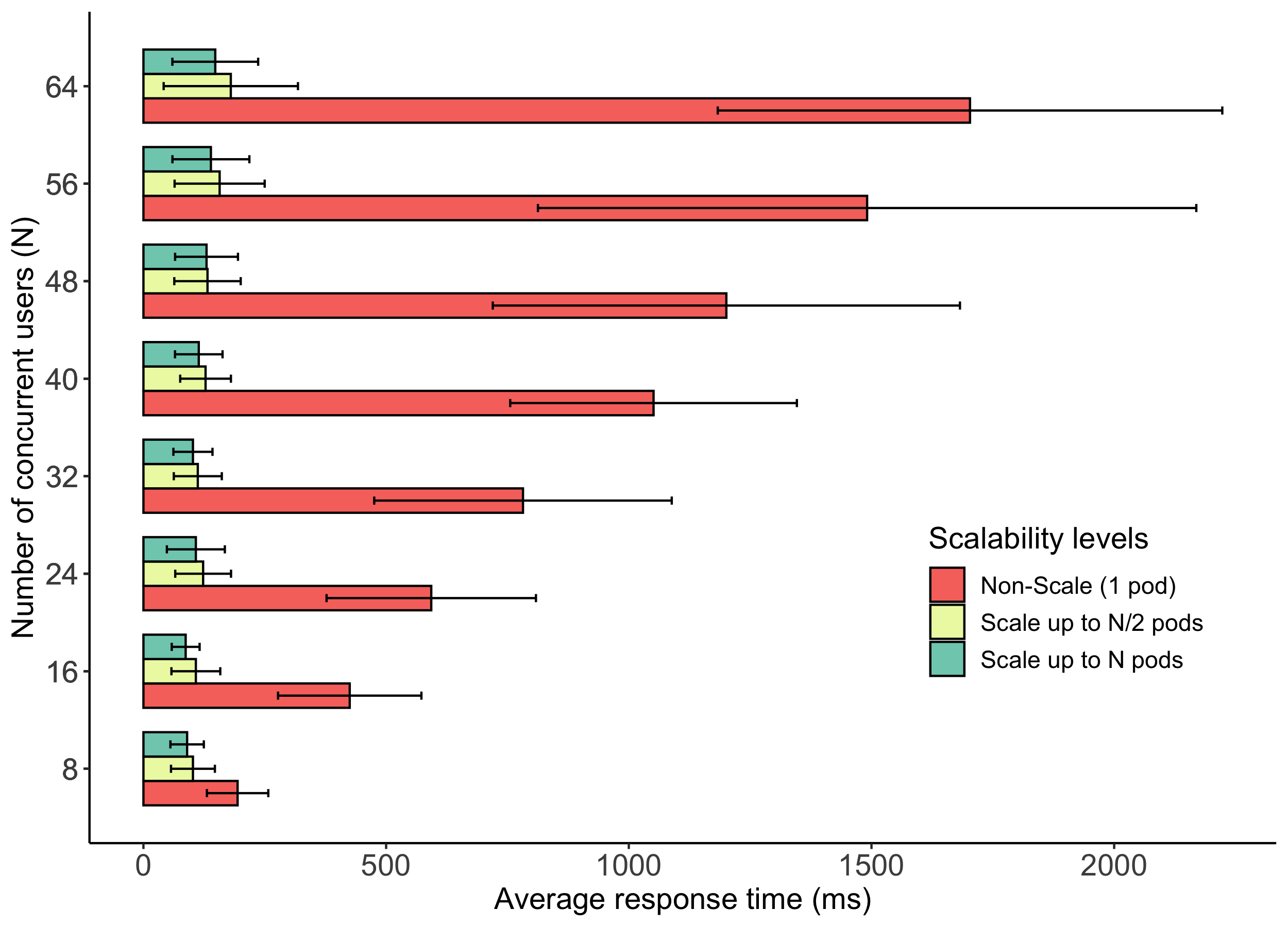}
    \caption{Scalability evaluation on Qiskit QRNG function.}
    \label{fig:scalable-eval}
\end{figure}

Figure \ref{fig:scalable-eval} demonstrates the result of our benchmarking. Overall, it is clear that if the function is non-scalable, the average response times for high-demand scenarios significantly increase. The previous section shows that the average response time for the 10-qubit Qiskit QRNG function is 81 ms. This figure jumps dramatically, up to 1703 ms, if 64 users use the function simultaneously. However, thanks to the containerization approach in our framework, we can quickly scale up deployment in seconds to ensure the response time is maintained. We can see that the average response time fluctuates between 87 to 148 ms if we scale up to N pods or from 102 to 180 ms when the number of pods is N/2.

\subsection{Case study 2: Shor’s algorithm with IBM Quantum Cloud Provider}
Shor’s algorithm \cite{shoralgo} is one of the most famous quantum algorithms for proving the advantage of quantum computing together with its classical counterpart. It is well-known for finding the prime factors of integers in polynomial time, which raises the severe risk for classical cryptography based on the security of large integers such as RSA. 

\subsubsection{Implement the Shor’s algorithm as a service using Qiskit API}
\begin{figure}[htb]
    \centering
    \includegraphics[scale=0.37]{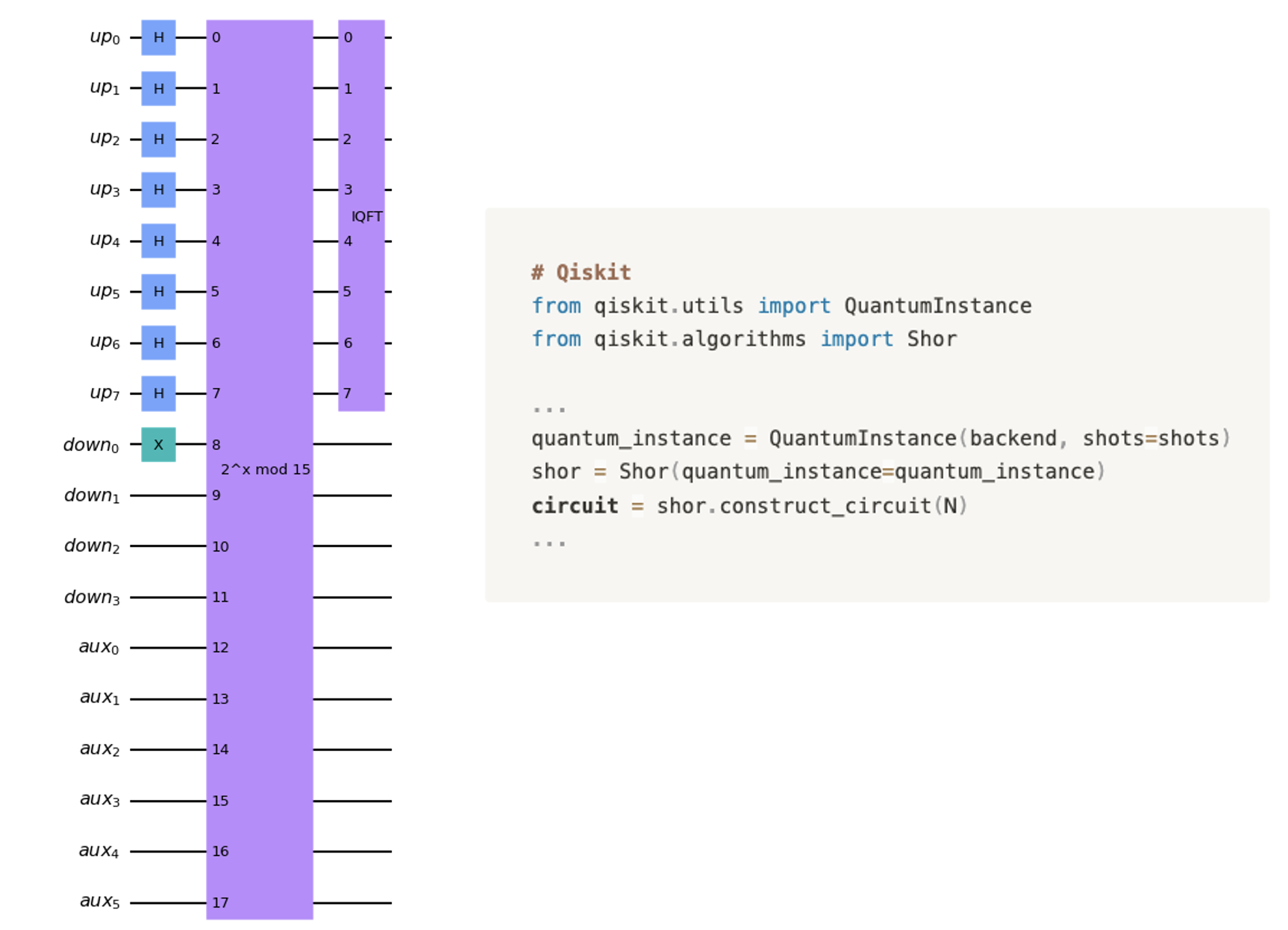}
    \caption{A simplified quantum circuit (left) and Qiskit code snippet (right) for implementing Shor algorithm to factorize 15}
    \label{fig:shor-circuit}
\end{figure}

In this case study, we demonstrate the implementation of Shor’s algorithm as a QFaaS function (Shor function) by utilizing the Qiskit Terra API \cite{qiskit}.
The quantum circuit for Shor’s algorithm is also dynamically generated based on the input number that end-users want to factorize. For example, to factorize 15, we need to use 18 qubits for the corresponding quantum circuit (Figure \ref{fig:shor-circuit}). Using the Shor class in \code{qiskit.algorithms}\footnote{https://qiskit.org/documentation/stubs/qiskit.algorithms.Shor.html}, we need to define a simple code to generate the quantum circuit to factorize the integer N, then execute that circuit at the appropriate backend device selected by QFaaS or through the end-users.

\subsubsection{Deploy and invoke Shor function}
The deployment process of the Shor function is similar to the QRNG function in the previous section.

For example, we use the following request and submit it to the Shor function, using the 27-qubit quantum computer (\code{ibm_cairo} node) at IBM Quantum Provider \cite{ibmq} for factorizing 15 (Code \ref{lst:shor-req}). After finishing the execution, the response data is as the following sample at Code \ref{lst:shor-rep}. We got the prime factors of 15 are 3 and 5 as expected. Due to the limitation of the current NISQ devices, we keep these example data small to demonstrate the viable workflow of developing, deploying, and using functions at QFaaS. We aim to develop a scalable QFaaS framework that could handle large-scale algorithms relevant to practical applications.

\begin{minipage}{.5\textwidth}
\hfill
\begin{lstlisting}[language=json, caption=Sample request data of the Shor function to factorize 15, label={lst:shor-req}]
{
    "input": 15,
    "provider": "ibmq",
    "shots": 100,
    "wait_for_result": true,
    "backend_info": {
        "hub": "ibm-q-melbourne",
        "api_token": "",
        "device": "ibm_cairo",
        "autoselect": false }
}\end{lstlisting}
\end{minipage}\hfill
\begin{minipage}{.45\textwidth}
\begin{lstlisting}[language=json, caption=Sample response data of the Shor function, label={lst:shor-rep}]
{
    "result": [[3, 5 ]],
    "device": "ibm_cairo",
    "detail": {
        "required_qubits": 18,
        "shots": 100 
        }
}\end{lstlisting}
\end{minipage}
\subsubsection{Performance Evaluation}
In this evaluation, we compare the actual performance of the Shor function with today's quantum computers provided by IBM Quantum \cite{ibmq}. We pick five adequate integer numbers for the test cases, including 15, 21, 35, 39, and 55. All experiments are conducted on a 27-qubit quantum computer (\code{ibm_cairo}, using Falcon r5.11 quantum processor) and the QASM simulator (\code{ibmq_qasm_simulator}). 

\begin{figure}[htbp]
    \centering
    \includegraphics[scale=0.13]{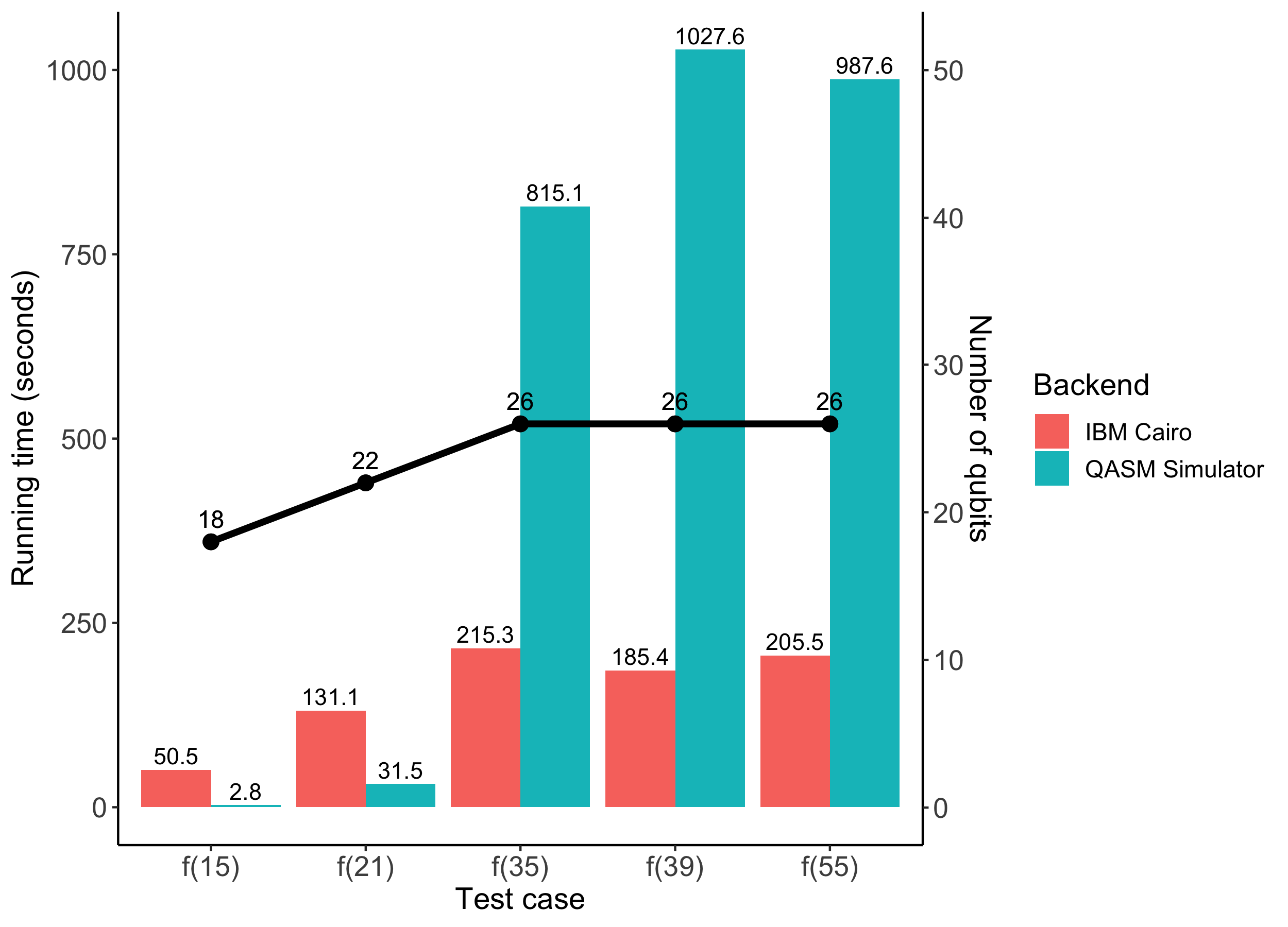}
    \caption{Performance evaluation of Shor function on (\code{ibm_cairo}) physical quantum backend and QASM simulator provided by IBM Quantum. f(15), f(21), f(35), f(39), and f(55) are five implemented test cases to factorize 15, 21, 35, 39, and 55, respectively.}
    \label{fig:shor-eval}
\end{figure}

Every time we invoke the Shor function with each test case, an appropriate circuit (quantum job) will be generated and sent to the IBM Quantum provider. Then, each quantum job will be validated and kept in the queue (from seconds to hours) before being executed in the backend due to the current fair-share nature of the IBM Quantum services. Therefore, to make a fair comparison, we only measure the actual running time (including the circuit validation and running in the system time, without the queuing time). We also execute each quantum job 100 times (shots) to ensure that the final result of all factorization problems is correct. In Figure \ref{fig:shor-eval}, the bar chart illustrates the actual running time, and the line chart indicates the number of qubits used for each test case in both backend devices. These input numbers need less than 27 qubits to build a corresponding circuit. Regarding the run time, we can see that the QASM simulator is much faster than the quantum computer when the number of required qubits is small, from 18 to 22 (for factorizing 15 and 21). However, we can see the opposite trend when executing 26-qubit circuits to factorize 35, 39, and 55. These circuits cost around 3 minutes to complete in an IBM Cairo quantum computer, whereas the QASM simulator takes 13.5 to 17 minutes to finish the execution. A significant reason for the considerable delay of the QASM Simulator in these test cases could be the complexity of the 26-qubit quantum circuit for the Shor algorithm, which requires a lot of resources to simulate. These results could give us insight into the selection order of current quantum computing services for developing quantum software. We could use the quantum simulator for prototyping and testing phases before entering the production stage with the quantum computers.

\section{Discussion}
By designing and developing QFaaS, we found that combining quantum and classical tasks along with the serverless-based model is possible and even an effective way to deal with the capabilities of current NISQ devices. Our framework provides a seamless quantum software development environment that allows software engineers to quickly develop, build, deploy, and eventually offer their quantum functions to end-users. End-users can also integrate these quantum functions into their existing classical application, especially suitable for microservice applications.

Inspired by the advantages of the Serverless and DevOps models, combined with the Quantum Cloud Computing paradigm, we have designed and implemented QFaaS with a set of essential features for creating a unified environment for developing quantum applications:

\begin{itemize}
    \item \textbf{Multiple quantum SDKs and languages supported}: QFaaS supports four quantum SDKs, and widely popular languages, including Qiskit \cite{qiskit}, Cirq \cite{cirq}, Q\# \cite{qsharp}, and Braket \cite{amz-braket-sdk}. 
    \item \textbf{Containerized quantum environment}: The engineers can develop quantum functions without any concerns about environment setup. The function will be containerized into an appropriate Docker image, including all necessary libraries defined by quantum software engineers. This approach makes the deployment more flexible and allows quantum developers to easily migrate their functions to other systems.
    \item \textbf{User-friendly Web UI with built-in IDE}: Quantum engineers can easily create, update, delete, manage and monitor their quantum function by using QFaaS Dashboard. We also integrate a built-in IDE in the QFaaS Dashboard, allowing quantum software engineers to write and update their quantum codes directly.
    \item \textbf{Local software development environment with CLI tool}: Our platform also supports using local IDE such as Visual Code for the development process. Then, they can upload the function codes to QFaaS through the CLI tool with all actions they can do through the Dashboard.
    \item \textbf{Hybrid quantum-classical functions}: Quantum functions could include both quantum and classical parts (using Python), which supports hybrid computation. Quantum parts can be run on multiple quantum simulators (QASM simulator by Qiskit or built-in simulator by Cirq, Q\#, and Braket) or external quantum providers (supported both IBM Quantum \cite{ibmq} and Amazon Braket \cite{amazonbraket}). Classical parts (pre-processing, post-processing, or other classical processing) can be run on classical computation nodes at the Kubernetes cluster, where quantum functions are deployed.
    \item \textbf{Quantum API gateway}: After deploying, quantum functions will be published for end-users using an API gateway to use and integrate into their existing microservice application. Each function has a unique URL that allows the end-user to invoke or incorporate into other existing microservice applications as an API.
    \item \textbf{Continuous Integration and Continuous Deployment (CI/CD) pipelines}: With the DevOps-oriented approach, QFaaS creates a seamless software development process that continuously delivers value to end-users. After finishing the development process, quantum functions will be automatically compiled, containerized, and deployed to the Kubernetes cluster. When the engineer updates a quantum function, it will also automatically deploy and ensure the updated services are successfully deployed before terminating the old one to guarantee the continuous experience at the end-users side.
    \item \textbf{No vendor lock-in}: The serverless feature of QFaaS is built on top of OpenFaaS \cite{openfaas} and Kubernetes, both of which are well-known classical platforms. This way, we could provide deployment flexibility and avoid the vendor lock-in problem, i.e., it could be deployed in any cloud cluster. We also demonstrate the ability to use external quantum providers, such as IBM Quantum, to execute quantum tasks and plan to extend QFaaS capabilities to support other providers when they are accessible in our region.     
    \item \textbf{High scalability and auto-scaling}: Ensures high availability and scalability for future expansion by deploying functions on the Kubernetes cluster. QFaaS also leverages the advantages of Kubernetes to support auto-scaling features, i.e., automatically scaling up or down vertically to adapt to the number of user requests.
    \item \textbf{Monitor function and system status}: The engineer or system manager can monitor the system status and all deployed functions using built-in monitoring components in QFaaS.
\end{itemize}

These features ease the quantum software development burden, enabling software engineers to focus on developing more complex quantum applications to achieve quantum advantages. Our work contributes to bridging the gaps between classical computing and the future computing model for developing practical quantum applications.

\section{Conclusions and Future Work}
In this work, we have developed QFaaS - a unified framework for developing quantum function as a service, enabling traditional software engineers to leverage their knowledge and experience to adapt to quantum counterparts in the \textit{Noisy Intermediate-Scale Quantum} era quickly. Our framework integrates several state-of-the-art methods such as containerization, DevOps, and the serverless model to reduce the burden for quantum software development and pave the way toward combining hybrid quantum and classical components. QFaaS provides essential features with multiple quantum software environments, leveraging the well-known quantum SDKs and languages to develop quantum functions running on quantum simulators or even a physical quantum computer. The current implementation of QFaaS demonstrates the possibility and advantages of our framework in developing quantum function as a service and continuously bringing its value to end-users.

Due to the current limitation around access to quantum cloud services from our region, we are able to demonstrate the experiments with IBM Quantum and Amazon Braket at the moment. We plan to extend QFaaS's ability to connect with other providers in the future. We also are developing a machine learning (ML) based approach for automatic selection of quantum backend for hybrid quantum-classical applications. We will enhance the security and scalability capabilities in QFaaS to support a large number of requests from multiple users. As Quantum Software Engineering is still an emerging area of research with numerous challenges, there is a need for significant research effort to make it reliable and simultaneously adapt to rapid advances in quantum hardware.

\textbf{Software Availability:}
We will release the QFaaS framework in open source to develop our strong collaboration in building a unified quantum serverless framework and contribute positively to making the quantum transition in software development smooth and seamless. It will be available at https://github.com/cloudslab/


\begin{acks}
This work is partially supported by the University of Melbourne by establishing an IBM Quantum Network Hub and the Nectar Research Cloud (offered by Australian Research Data Commons) at the University. We appreciate the support from Strangeworks Backstage Program for providing access to Amazon Braket quantum service. Hoa T. Nguyen acknowledges the support from the Science and Technology Scholarship Program for Overseas Study for Master’s and Doctoral Degrees, Vin University, Vingroup, Vietnam.

\end{acks}

\bibliographystyle{ACM-Reference-Format}
\bibliography{references}

\appendix









\end{document}